\documentclass[twocolumn]{aastex62}
\usepackage{amsmath}
\usepackage[flushleft]{threeparttable} 
\usepackage{array}
\newcommand{\nTOri}{10} 
\newcommand{\nVbig}{39}
\newcommand{\nVsmall}{40}
\newcommand{\nLP}{11}
\newcommand{\nrobeam}{$22\arcsec$}
\newcommand{\resolutionpc}{$0.04$~pc}
\newcommand{\distance}{$414$~pc}
\newcommand{\Msun}{$M_\odot$}
\newcommand{\Mdot}{$M_\odot$~yr$^{-1}$}
\newcommand{\kms}{km~s$^{-1}$}
\newcommand{\co}[1][]{\ensuremath{^{#1}}CO}

\shorttitle{Shells in Orion A}
\shortauthors{Feddersen {\em et al.}}

\begin{document}

\title{Expanding CO Shells in the Orion A Molecular Cloud}

\author{Jesse R. Feddersen}
\affiliation{Department of Astronomy, Yale University, New Haven, CT 06511, USA}
\author{H\'ector G. Arce}
\affiliation{Department of Astronomy, Yale University, New Haven, CT 06511, USA}
\author{Shuo Kong}
\affiliation{Department of Astronomy, Yale University, New Haven, CT 06511, USA}
\author{Yoshito Shimajiri}
\affiliation{Laboratoire AIM, CEA/DSM-CNRS-Universit$\acute{\rm e}$ Paris Diderot, IRFU/Service d'Astrophysique, CEA Saclay, F-91191 Gif-sur-Yvette, France}
\author{Fumitaka Nakamura}
\affiliation{National Astronomical Observatory of Japan, 2-21-1 Osawa, Mitaka, Tokyo 181-8588, Japan}
\author{Chihomi Hara}
\affiliation{Department of Astronomy, The University of Tokyo, 7-3-1 Hongo Bunkyo, Tokyo 113-0033, Japan}
\author{Shun Ishii}
\affiliation{Joint ALMA Observatory, Alonso de C\'ordova 3107 Vitacura, Santiago, Chile}
\author{Kazushige Sasaki}
\affiliation{Department of Physics, Niigata University, 8050 Ikarashi-2, Niigata 950-2181, Japan}
\author{Ryohei Kawabe}
\affiliation{National Astronomical Observatory of Japan, 2-21-1 Osawa, Mitaka, Tokyo 181-8588, Japan}

\email{jesse.feddersen@yale.edu}

\accepted{June 4, 2018}
\submitjournal{ApJ}

\begin{abstract}
We present the discovery of expanding spherical shells around low to intermediate-mass young stars in the Orion A giant molecular cloud using observations of \co[12](1-0) and \co[13](1-0) from the Nobeyama Radio Observatory 45-meter telescope. The shells have radii from 0.05 to 0.85~pc and expand outward at 0.8 to 5~\kms. The total energy in the expanding shells is comparable to protostellar outflows in the region. Together, shells and outflows inject enough energy and momentum to maintain the cloud turbulence. The mass-loss rates required to power the observed shells are two to three orders of magnitude higher than predicted for line-driven stellar winds from intermediate-mass stars. This discrepancy may be resolved by invoking accretion-driven wind variability. We describe in detail several shells in this paper and present the full sample in the online journal.
\end{abstract}

\keywords{ISM: bubbles --- ISM: clouds --- ISM: individual objects (Orion A) ---  stars: formation --- stars: pre-main sequence --- stars: winds, outflows}

\section{Introduction}
Stars form via the gravitational collapse of molecular gas in the densest parts of giant molecular clouds (GMCs) \citep{McKee07,Dunham14}. The efficiency of star formation observed in GMCs in the Milky Way is much lower than expected if gravity is the only force at work. The low star-formation efficiency has been attributed to magnetic fields \citep{Mestel56,Shu83,Crutcher12}, short GMC lifetimes \citep{Murray11,Dobbs13}, and turbulence \citep{Larson81,Mac-Low04,Federrath15}.

GMCs are turbulent, characterized by a log-normal column density probability distribution function (\citealp{Vazquez-Semadeni94};~c.f.~\citealp{Alves17}) and logarithmic relationship between their physical size and velocity width \citep{Larson81}. However, turbulence in GMCs rapidly decays within a cloud crossing time \citep{Mac-Low98,Stone98,Padoan99}. If turbulence is responsible for supporting clouds, it must be maintained by some mechanism.

Mechanical and thermal feedback from forming stars can deposit significant energy and momentum into GMCs. This can help maintain cloud turbulence and support against gravitational collapse, helping to explain the low star-formation efficiencies observed in GMCs \citep{Nakamura07,Federrath15}. It remains uncertain how these mechanisms maintain cloud turbulence. Therefore, it is important to measure how much energy and momentum are supplied by different stellar feedback mechanisms.

Young protostars launch accretion-driven collimated outflows \citep{Arce07,Frank14,Bally16}. More evolved pre-main sequence and main sequence stars are less embedded than their younger counterparts and drive wide-angle or spherical winds \citep{Castor75,Vink00,Bally11}. 

Massive stars have long been known to drive powerful stellar winds that impact the surrounding interstellar medium. In the last several years, \emph{Spitzer} surveys of the galactic plane have revealed `bubbles', mostly powered by massive stellar winds \citep{Churchwell06,Churchwell07,Beaumont10,Deharveng10,Beaumont14}.

\citet{Arce11} discovered expanding shells in the Perseus molecular cloud, which is not forming massive 
ionizing stars. They showed that these expanding shells have enough energy and momentum to drive cloud turbulence in Perseus.
These shells must be driven by intermediate-mass stars or protostars. 
\citet{Offner15} found that a spherical stellar wind of sufficient strength can drive Perseus-like shells when placed in a simulated turbulent cloud. In the Taurus molecular cloud, another low-mass star forming region, \citet{Li15} identified many expanding shells. 

We identify expanding spherical structures of molecular gas in the Orion A GMC, hereafter called `shells'. These shells are similar to the structures first found in the Perseus Molecular Cloud by~\citet{Arce11} and later in the Taurus Molecular Cloud by~\citet{Li15}. In Orion, shell-like structures have been identified by \citet{Heyer92} and \cite{Nakamura12}. This study is the first systematic search for expanding shells in Orion.

The Orion A GMC, located behind the Trapezium OB association, is the nearest massive star forming region. Orion A has been extensively observed at all wavelengths, including CO spectral mapping by \citet{Bally87}, \citet{Wilson05}, \citet{Shimajiri11}, \cite{Ripple13}, and \cite{Berne14}. The cloud is filamentary and exhibits a North-South velocity gradient of about 9~\kms~\citep{Bally08}. The cloud is forming both massive stars, traced by the HII regions M42 and M43 in the north, as well as lower mass stars along the `integral shaped filament' and in the NGC 1999 and L1641 clusters in the southern part of the cloud. We adopt a distance to Orion A of 414~pc \citep{Menten07}.

In Section~\ref{sec:methods} we describe our data and how we find and characterize shells. In Section~\ref{sec:results} we present the shells found in Orion A and discuss several shells in detail. In Section~\ref{sec:impact}, we discuss the mass, momentum, and kinetic energy of the shells. In Section~\ref{sec:discussion} we compare the impact of the shells on the cloud to turbulence and protostellar outflows and discuss mechanisms that may drive the shells.

\section{Methods}\label{sec:methods}

\subsection{Nobeyama Radio Observatory 45m Observations}\label{sec:nro} We briefly describe the observations here. For more detail, see~\citet[Section 2.2]{Kong18}. From 2007 to 2017, we carried out observations of \co[12](1-0, 115.271~GHz) and \co[13](1-0, 110.201~GHz) in Orion A with the Nobeyama Radio Observatory 45-meter telescope (NRO). From 2007 to 2009 and 2013 to 2014, we used the 25-beam BEARS focal plane array. With BEARS, we used 25 sets of 1024 channel auto-correlators with a 32 MHz bandwidth for a velocity resolution of $\sim0.1$~\kms~at 115~GHz \citep{Shimajiri11,Nakamura12,Shimajiri14}. From 2014 to 2017, we used the new 4-beam FOREST receiver with the SAM45 spectrometer for a velocity resolution of $\sim0.04$~\kms~at 115~GHz.

We combine the FOREST and BEARS maps for the best sensitivity and coverage. The final NRO map has a beam FWHM of $\sim$\nrobeam~(\resolutionpc~at a distance of~\distance) and a velocity resolution of $\sim0.22$~\kms.

\subsection{Infrared Data}\label{sec:irdata}
To assist with our search for expanding shells (see Section~\ref{sec:criteria}), we use archival infrared images from the Spitzer Heritage Archive. We search IRAC 3.6/8~$\mu$m and MIPS 24~$\mu$m for dust rings correlated with the CO shells. IRAC images are from Spitzer Programs 43 and 30641 (PI: Fazio). MIPS images are from Spitzer Programs 47 and 30641 (PI: Fazio). We also look for correlated structures in the effective dust temperature maps from \citet{Lombardi14} produced by fitting the spectral energy distribution (SED) of the \emph{Herschel} and \emph{Planck} maps.

\subsection{Source Catalogs}\label{sec:catalogs}
To match expanding shells with the stars that may be driving them, we use catalogs of intermediate-mass stars and young stellar objects (YSOs) in Orion A. We queried \emph{Simbad} for all stars with spectral type B, A, or F in the area. These intermediate-mass main sequence and pre-main sequence stars are good candidates for driving CO shells.

We also use the Spitzer Orion catalog of protostars and pre-main sequence stars produced by \citet{Megeath12}. The stars are classified as protostars or disk stars (pre-main sequence stars) by their infrared photometry. Stars with rising or flat SEDs between 4.5 and 24~$\mu$m are classified as protostars. All other stars with infrared excess are considered to have disks and have dispersed their natal envelope. These (mostly low-mass) young stars are potential driving sources for shells, especially when clustered \citep[see][]{Nakamura12}.

\subsection{Identifying shells}\label{sec:criteria}
We identify shell candidates visually by searching in the CO channel maps for circular cavities that change in size with velocity - a signature of expansion. We also look in the position-velocity (PV) diagram for a $\cup$~or~$\cap$-shaped feature indicating expansion \citep[see][Figure 5]{Arce11}.

We match the shell candidates against the source catalogs described in Section~\ref{sec:catalogs} to identify stars that may drive the shells. If a YSO from the Spitzer Orion catalog or an intermediate-mass BAF-type star is located inside the shell radius in projection, we consider this a potential driving source of the shell. The source need not be at the center of the shell. If the driving mechanism is not continuous, we may expect a star to have moved from the shell center. \citet{Hartmann02} found an average relative velocity of 0.2~\kms~between protostars and gas in the Taurus Molecular Cloud. In Orion, \citet{Tobin09} found a similar velocity difference between stars and gas. In 1-2~Myr, a source moving at 0.2~\kms~may travel 0.2-0.4~pc (100-200\arcsec) from the center of the shell. This distance is similar to the typical radius of a shell (Table~\ref{tab:shells}).

We use infrared images of dust emission to identify dust swept up in expanding shells. Using the Spitzer IRAC and MIPS maps described in Section~\ref{sec:irdata}, we search for rings of dust emission that are correlated with CO shells. Using the Planck-Herschel map, we search for dust temperature correlations with the shells.

We score the reliability of each shell candidate by the number of criteria it satisfies. The criteria used to score each shell are:
\begin{enumerate}
\item The CO channel maps show expanding velocity structure.
\item The position-velocity diagram of the shell shows an expansion signature ($\cup$ or $\cap$ shape) as modeled in Section~\ref{sec:model}.
\item The shell has a circular shape in integrated CO and/or IR dust emission. To satisfy this criterion, the shell emission must be visible around at least half of the circular cavity.
\item The CO shell is correlated with infrared nebulosity in at least one band. This criterion is satisfied if any part of the observed CO shell (including a central cavity) is traced by a similar feature in an infrared band.
\item The shell contains a candidate driving source.
\end{enumerate}

These criteria are subjective, and are not intended to definitively determine which shells are "real" but to give a relative measure of significance. We encourage readers to use the included figures to judge these criteria for themselves.

\subsection{Characterizing Shells}
We characterize each shell with four parameters: radius, thickness, expansion velocity, and central velocity. To find the most likely parameters, we use the model described below.
\subsubsection{A Simple Expanding Shell Model}\label{sec:model}
We use a simple model for an expanding shell adapted from~\cite{Arce11}. The model assumes uniform expansion, spherical symmetry and optically thin emission.

To create a model spectral cube of an expanding shell, we first randomly sample points from a spherical shell of uniform volume density with radius~$R$ and thickness~$dr$. The number of points we sample is chosen to ensure there are several points per resolution element of the final spectral cube. We assign each sample a line-of-sight velocity $v_z$ which scales with its displacement along the line-of-sight $z$ and radial displacement $r$ from shell center:
\begin{equation}\label{eq:model}
v_z = v_{\rm exp} \frac{z}{r} + v_0
\end{equation}
where $v_{\rm exp}$ is the expansion velocity of the shell and $v_0$ is the central velocity of the shell. We bin the sampled points by position on the sky and line-of-sight velocity to make a synthetic position-position-velocity cube with the same dimensions as the observed CO cube.\footnote{The model cube is padded by 5 pixels on each side and by 5 channels blueward and redward of the most extreme shell velocities. All velocities in this paper are taken with respect to the local standard of rest (LSR).}

The shell model described in~\cite{Arce11} can incorporate a turbulent cloud of uniform mean density. However, since we do not attempt to describe the underlying cloud properties, we simplify the model by removing the cloud component and only considering~$R$, $dr_{\textrm{shell}}$, $v_{\textrm{exp}}$, and the central velocity of the shell $v_0$.

We vary these four model parameters and visually compare the model and observed integrated emission and position-velocity diagrams of each shell candidate. Table~\ref{tab:shells} lists the parameters of the model that most closely matches the observations. We vary each parameter individually while holding the others fixed to visually estimate the parameter uncertainties reported in Table~\ref{tab:shells}. Expansion velocity is the most uncertain parameter, as most shell candidates are not detected over their entire velocity range. Therefore, the estimated expansion velocity may be considered a lower limit.

The model is meant to be a very idealized version of an expanding shell. Real shells are not symmetric; they inherit the turbulent structure of the cloud emission. Unlike the model, most observed shells are not completely contained within the cloud. Our model also assumes optically thin emission, which is unrealistic for \co[12] (and possibly \co[13]) over much of the cloud. Because the model is not flexible enough to account for these complications, we do not attempt a statistical fit of the model to the CO data. The parameter ranges reported in Table~\ref{tab:shells} produce the range of models that most closely resemble the observed shells.

Model PV diagrams are shown in Section~\ref{sec:results}. These figures show that matching any one model to an observed shell is difficult and this is reflected in the uncertainties on the model parameters we report in Table~\ref{tab:shells}.

\begin{figure*}[htb!]
\centering
\includegraphics[]
{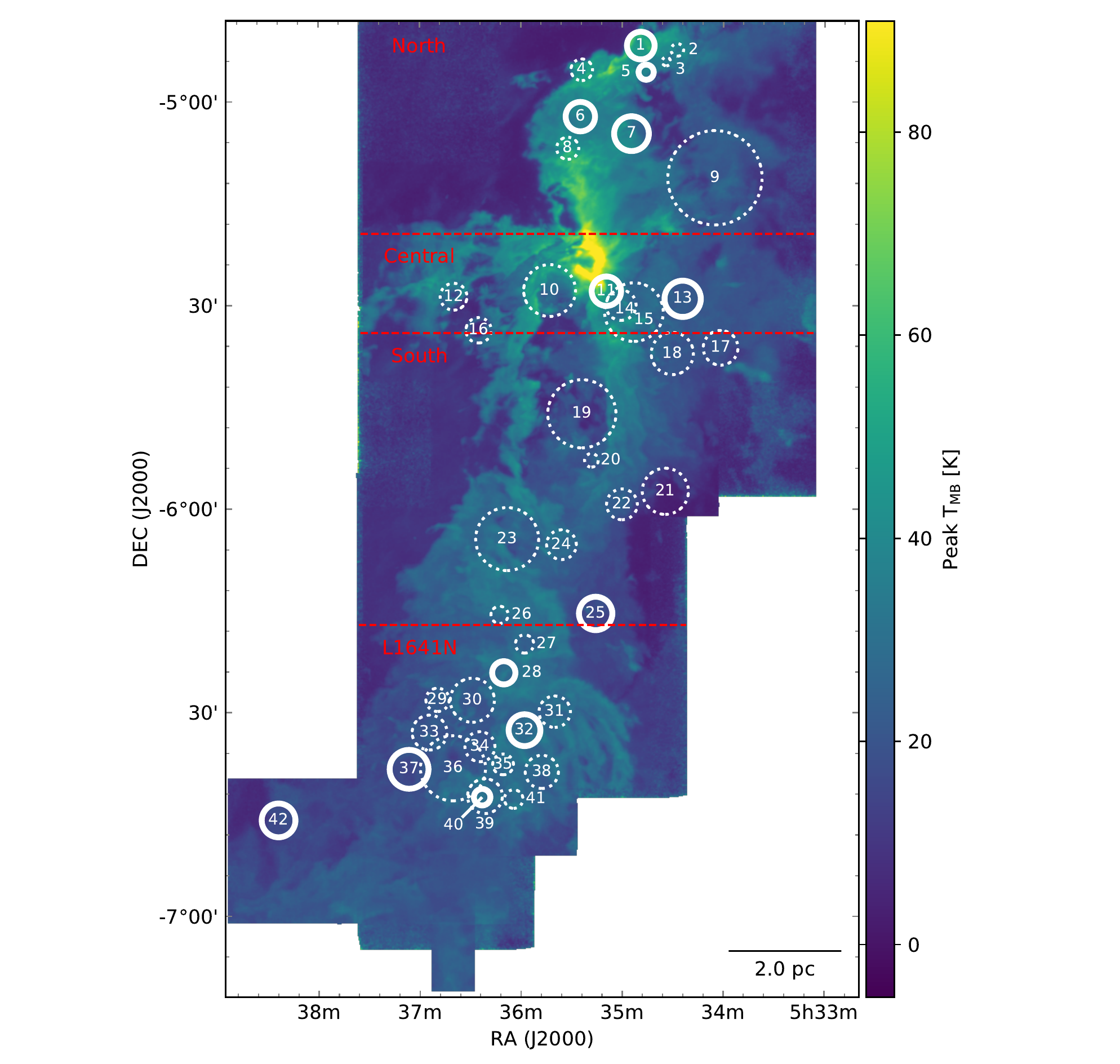}
\caption{Peak \co[12] intensity with shell candidates. Solid circles indicate 12 shells that satisfy all criteria listed in Section~\ref{sec:criteria}. Dashed circles show 30 less robust shell candidates.}
\label{fig:12co_peak_shells}
\end{figure*}

\section{Results}\label{sec:results}
We identify 42 shell candidates in Orion A. Figure~\ref{fig:12co_peak_shells} shows the peak \co[12] brightness temperature in Orion A with shell candidates overlain. Table~\ref{tab:shells} lists the estimated range in model parameters (radius, thickness, expansion velocity, and systemic velocity) of the shell candidates. 

Table~\ref{tab:criteria} lists the criteria (defined in Section~\ref{sec:criteria}) each shell candidate satisfies. We assign a confidence score of 1 to 5 to each shell equal to the number of criteria the shell satisfies. A score of 1 means the shell candidate was identified in CO channel maps but satisfies no other criteria. A score of 5 is given to the shells which satisfy all criteria. The properties of this most reliable subset of shells do not differ systematically from the full set.

We present figures detailing all 42 shell candidates in the online journal. For each shell, we show a representative infrared image with integrated CO contours, CO channel maps\footnote{The figures include CO channels that show clear shell emission. Sometimes the best model central velocity listed in Table~\ref{tab:shells} corresponds to a channel that does not contain emission. In this case, the shell velocity range in Table~\ref{tab:shells} will not be the same as the velocity range shown in the channel maps.}, and a CO position velocity diagram. We discuss four shells in detail here. These four are not meant to be representative of the entire sample. They are chosen for their CO morphology and interesting candidate driving sources which show clear signs of intermediate-mass stellar feedback on the cloud.

\subsection{A Shell Near The Herbig Ae Star T Ori}\label{sec:shell_TOri}

\begin{figure*}
\centering
\includegraphics[width=13cm]{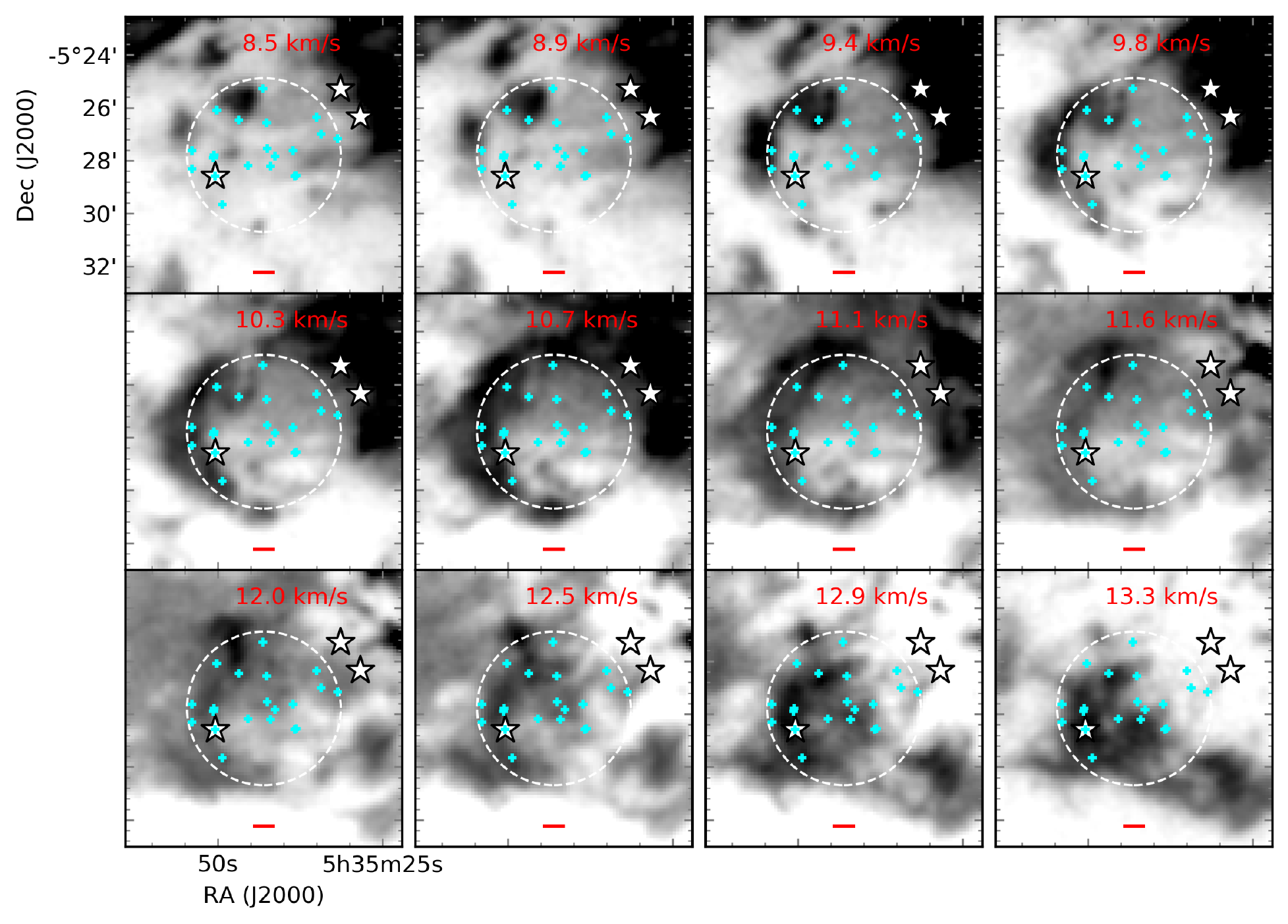}
\caption{\co[12] channel maps of Shell~\nTOri, thought to be powered by T Ori. Stars indicate intermediate-mass stars of spectral type B, A, or F and are labeled in Figure~\ref{fig:shell_TOri_ir}. Cyan crosses indicate pre-main sequence stars from the Spitzer Orion catalog. The full best-fit radius is shown as a dashed white circle. Velocities are with respect to the local standard of rest. The red scalebar has a length of 0.1~pc. The full figure set of channel maps (42 images) is available in the online journal.}
\label{fig:shell_TOri_channels}
\end{figure*}
Shell~\nTOri~is about 0.16~degrees~($1.2$~pc) southeast of the massive molecular core OMC~1. The shell meets 4 of the criteria listed in Section~\ref{sec:criteria}.

\paragraph{CO Channel Maps}
This shell, like most in the catalog, was first discovered by inspecting the \co[12] channel maps (Figure~\ref{fig:shell_TOri_channels}). The shell first appears as disconnected clumps at 8.5~\kms. At higher velocities, the shell gains prominence and is most clearly seen as the C-shaped structure at 10.7~\kms. The shell emission decreases in radius in subsequent channels as the cross section of the shell on the sky shrinks. At 12-13.3~\kms, an unrelated spur of \co[12]~appears to the southwest of the shell. This spur is part of the larger scale expansion driven into the molecular cloud by the M42 HII region. This expansion, identified by \citet{Loren79} and \citet{Heyer92}, can also be seen near Shell~\nLP~in Figure~\ref{fig:shell_LPOri_channels}.

\begin{figure*}
\centering
\includegraphics[width=10cm]{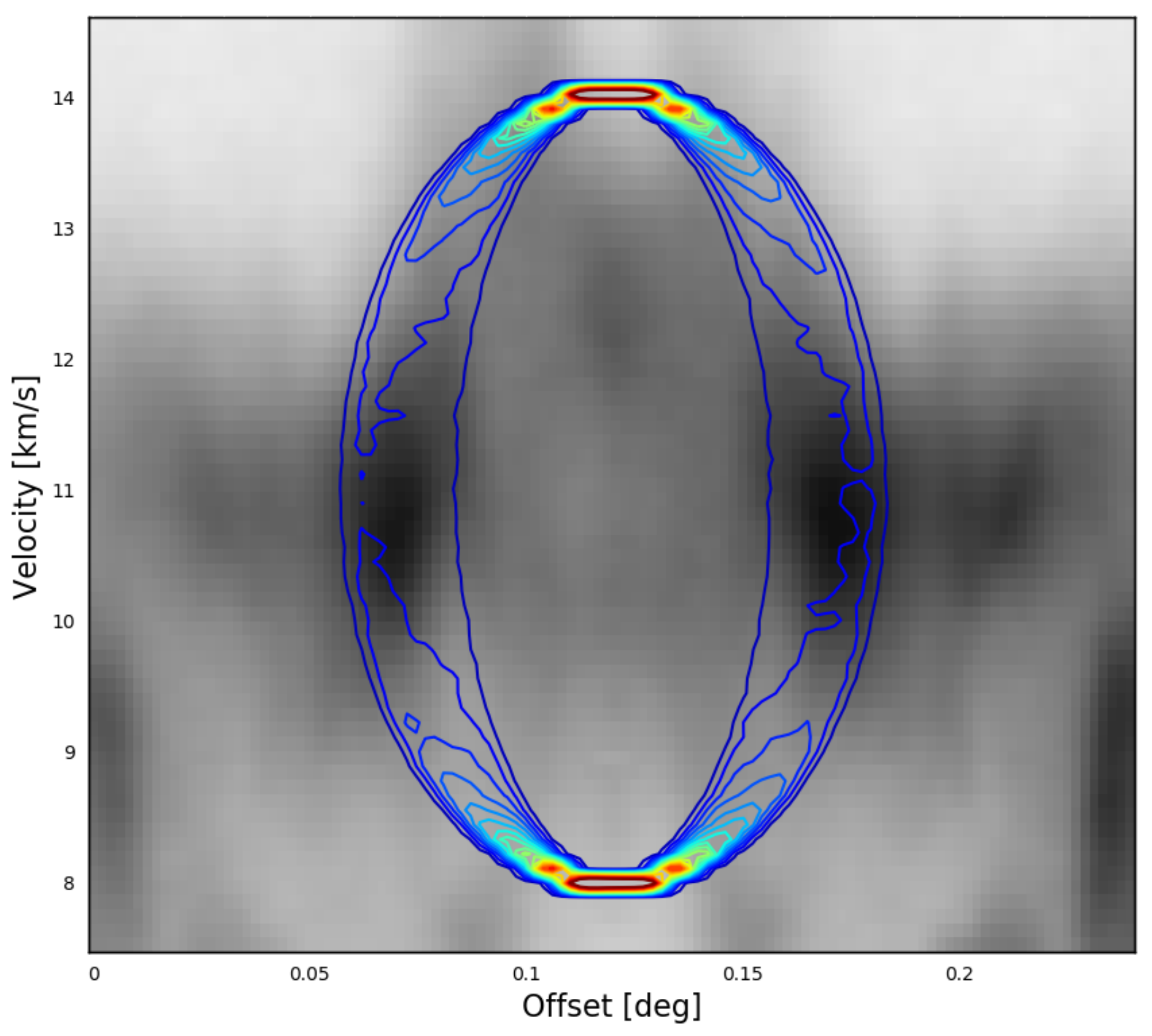}
\caption{Azimuthally averaged position-velocity diagram of \co[12] emission toward Shell~\nTOri. Darker colors indicate more intense emission. We extract emission along 4 equally spaced slices through the center of the shell and then average. The approaching and receding caps of the shell are not clearly detected. Contours show the model that best represents the shell. The model parameters are given in Table~\ref{tab:shells}. The full figure set of PV diagrams (42 images) is available in the online journal.}
\label{fig:shell_TOri_pv}
\end{figure*}

\paragraph{Position-Velocity Diagram}
Figure~\ref{fig:shell_TOri_pv} shows the position-velocity diagram of \co[12] across this shell. To increase the signal to noise in the PV diagram, we compute the azimuthally averaged PV diagram through the center of the shell at four equally spaced position angles. The PV diagram does not clearly show the $\cup$ or $\cap$-shaped signature expected of an expanding structure. However, averaging across many position angles may dilute the expansion signature if the shell is not azimuthally symmetric. In the case of Shell~\nTOri, the averaged PV diagram may dilute some of the emission at $v > 12.5$~\kms.

\paragraph{Infrared Nebulosity}

\begin{figure*}
\centering
\includegraphics[width=15cm]{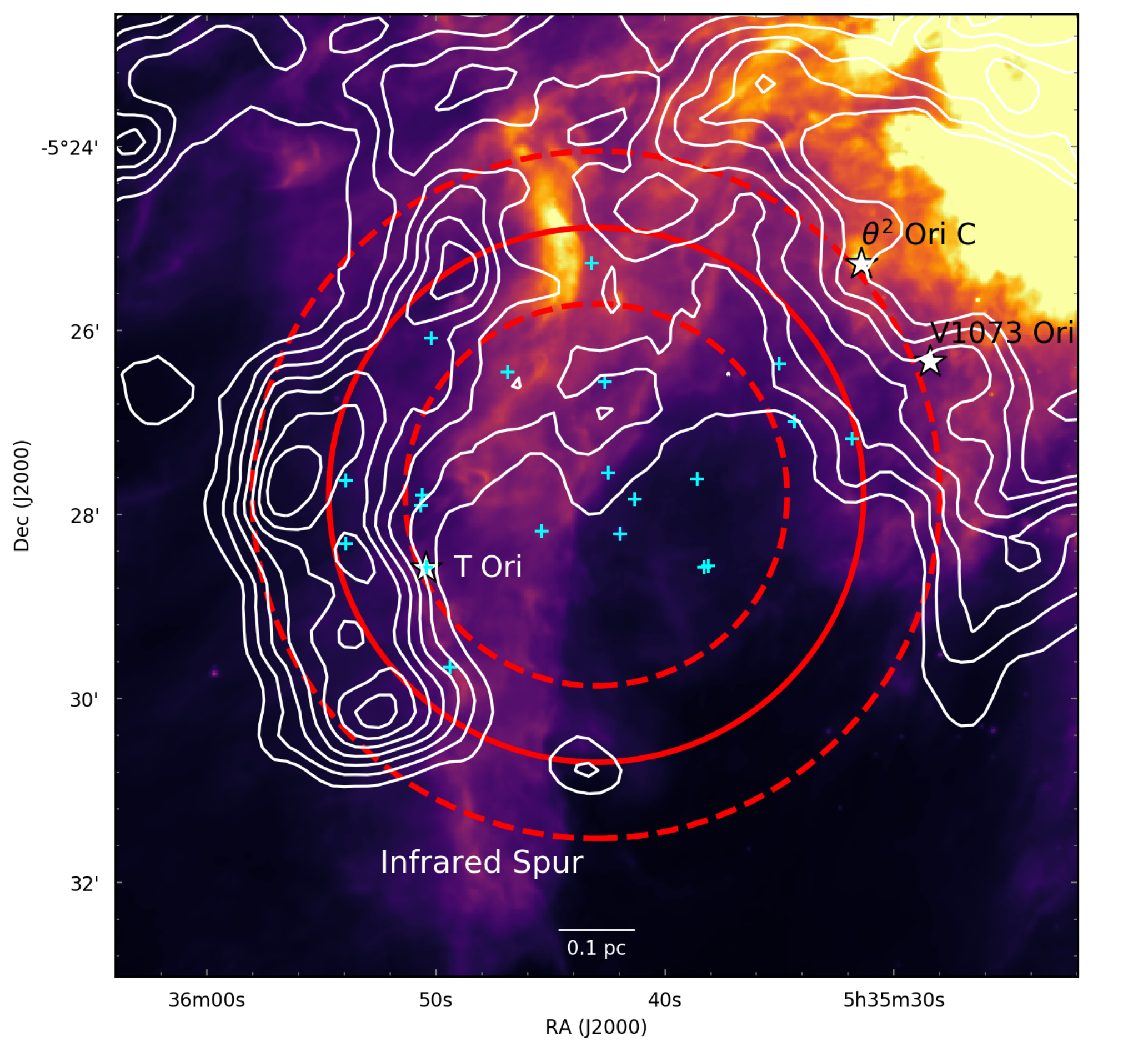}
\caption{Spitzer 8~$\mu$m map toward Shell~\nTOri. Contours show \co[13]~integrated from 8.5 to 13.5~\kms. Contours are drawn from 10 to 22$\sigma$ with steps of $2\sigma$, where $\sigma = 1.2$~K~\kms. Symbols are the same as Figure~\ref{fig:shell_TOri_channels}. The large solid circle and dashed annulus indicate the best-fit radius and thickness of the CO shell, respectively. The full figure set (42 images) is available in the online journal.}
\label{fig:shell_TOri_ir}
\end{figure*}

Figure~\ref{fig:shell_TOri_ir} shows the 8~$\mu$m map highlighting dust emission near the shell. The dust emission towards the west side of the shell is spatially coincident with the CO structure. An unrelated infrared-bright spur \citep[see][]{Shimajiri11,Shimajiri13} projected from north to south through the center of the shell highlights the cometary structure shaped by the hard ionizing radiation field from the Trapezium OB association to the northwest.

\paragraph{Potential Driving Sources}
This shell contains several intermediate-mass stars and protostars. T~Ori is a $5$~Myr old Herbig A2-3e star \citep{Hillenbrand92,Liu11} offset from the center of the shell by approximately $0.2$~pc to the southeast. \cite{Fuente02} identified  a cavity in integrated \co[13] and C$^{18}$O around T~Ori. They argue that intermediate-mass pre-main sequence stars like T~Ori excavate the molecular gas around them over time. They find the youngest stars in their sample at peaks of dense gas and more evolved pre-main sequence stars (like T Ori) in cavities, attributing this excavation to stellar winds. \cite{Liu11} modeled the spectral energy distribution of T~Ori, deriving an age of $\approx 5$~Myr and an accretion rate of $\approx 3 \times 10^{-7}$ \Mdot. Protostellar mass-loss rates are expected to be approximately 10-30\% of their accretion rates \citep[e.g.,][]{Pudritz07,Mohanty08}. T~Ori falls within the Herbig Ae/Be mass-loss rates of $10^{-8}$ to $10^{-7}$ \Mdot~measured by \citet{Skinner94}. The mass-loss rate required to power the shell around T~Ori is $\approx 10^{-6}$ \Mdot, an order of magnitude higher than the estimated mass-loss rate (See Table~\ref{tab:physics}; Section~\ref{sec:wind_energy}).

$\theta^{2}$Ori~C, located just outside the edge of the shell, is a B4/5 star in the Orion Nebula Cluster. Though it lacks spectral emission lines, this star has been included as a Herbig Be star by many authors based on its far-infrared excess \citep{The94}. \citet{Manoj02} argues that $\theta^{2}$Ori~C is a young ($\approx$ 1 Myr) pre-main sequence star surrounded by dust. X-ray observations show strong flares from this star, which \cite{Stelzer05} put forward as evidence for a low-mass T-Tauri companion to $\theta^{2}$Ori~C. \cite{Megeath12} classify $\theta^{2}$Ori~C as a pre-main sequence star with a disk, based on its mid-infrared colors.

V1073~Ori, located outside the edge of the shell, is an A0 star in the Orion Nebula Cluster \citep{Hillenbrand13} with an age of 5 Myr \citep{Hillenbrand97}. Because this star is at the same projected distance as $\theta^{2}$~Ori~C but much less massive, any impact on the shell from these two stars is likely dominated by $\theta^{2}$~Ori~C.

Another possibility is that this shell is shaped by the UV radiation field from the Trapezium cluster to the northwest. In this case, the shell could be seen as an extension of the cometary photon dominated region (PDR) to the south denoted the ``dark lane south filament" by \citet{Shimajiri11,Shimajiri13}.
However, the velocity of the PDR ranges from 5-8~\kms~while the shell is seen at 8-13~\kms. Thus, the shell is distinct in velocity-space from these cometary pillars.

Because of its proximity to the projected center of the shell and known winds, T~Ori is the most likely driving source of Shell~\nTOri.

\subsection{Two Nested Shells Around V380 Ori}\label{sec:shell_v380}
We identify two nested expanding shells near the young Herbig B9e star V380 Ori. Shell~\nVbig, the larger of the two, was first identified while searching the CO channel maps. The smaller Shell~\nVsmall~was found upon closer inspection for shells around potential driving sources. This region also contains several Herbig-Haro (HH) objects \citep{Stanke10} and CO outflows \citep{Morgan91,Moro-Martin99}.

\begin{figure*}
\centering
\includegraphics[width=13cm]{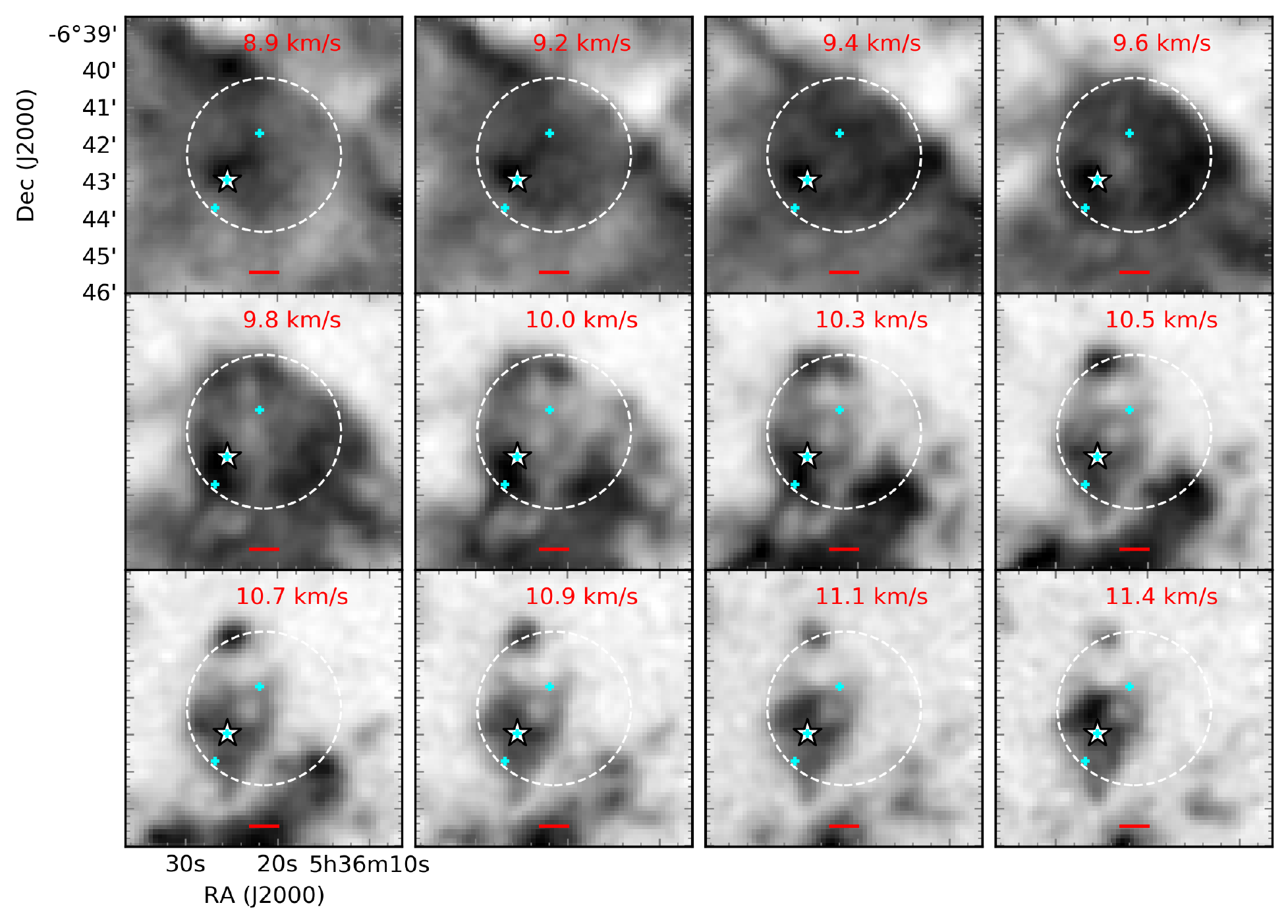}
\caption{\co[12] channel maps of Shell~\nVbig, the larger of the two shells around V380~Ori. Symbols are the same as Figure~\ref{fig:shell_TOri_channels}. The best-fit model radius is shown as a white dashed circle. The smaller Shell~\nVsmall~can be seen near V380~Ori (white star). The red scalebar has a length of 0.1~pc.}
\label{fig:shell_v380large_channels}
\end{figure*}

\begin{figure*}
\centering
\includegraphics[width=13cm]{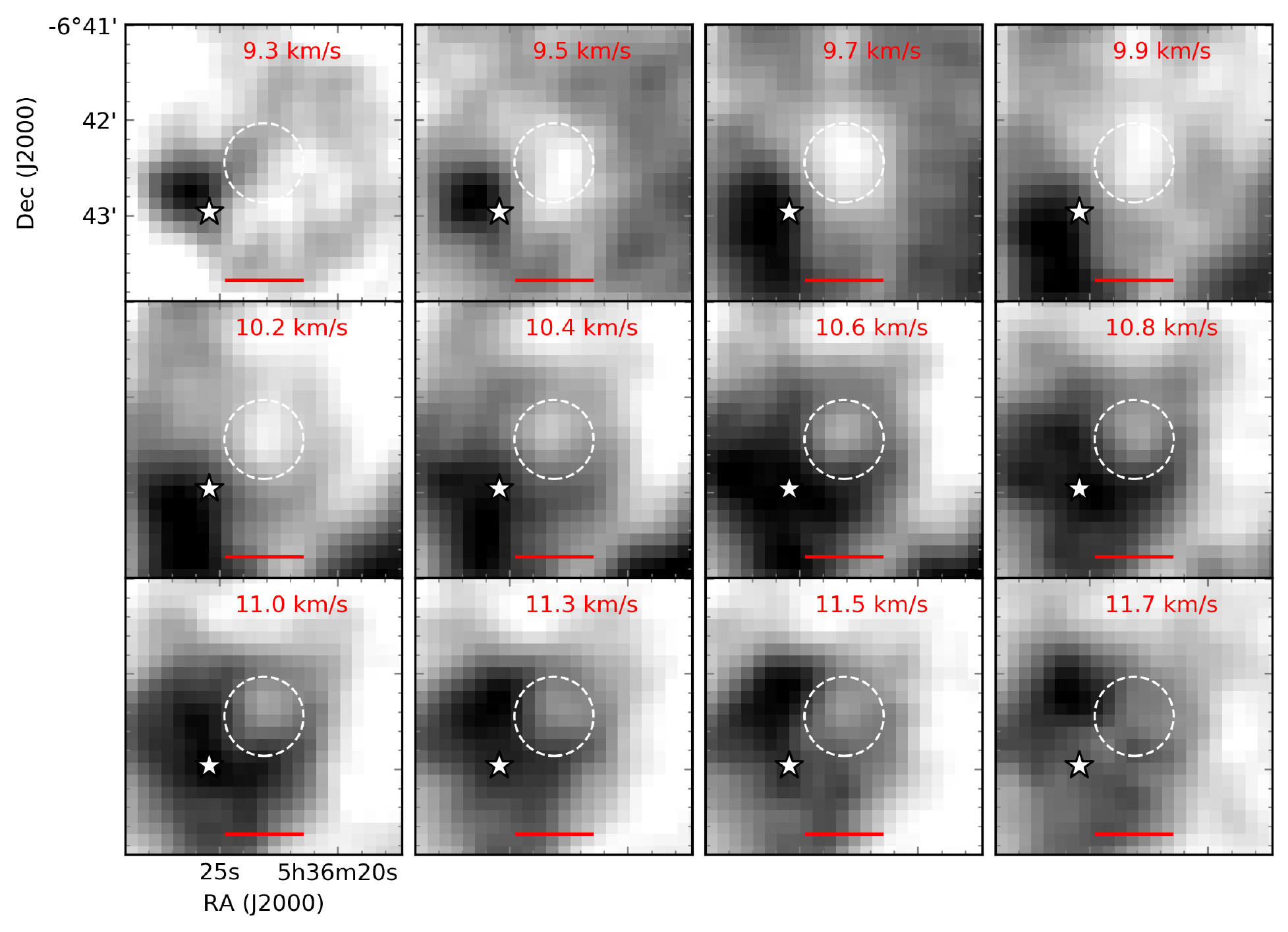}
\caption{\co[12] channel maps of Shell~\nVsmall, the smaller of the two shells associated with V380~Ori. Symbols are the same as Figure~\ref{fig:shell_TOri_channels}. The best-fit model radius is shown as a white dashed circle. The red scalebar has a length of 0.1~pc.}
\label{fig:shell_v380small_channels}
\end{figure*}

\paragraph{CO Channel Maps}
Figure~\ref{fig:shell_v380large_channels} shows Shell~\nVbig~in \co[12]. Shell~\nVbig~ is most clearly defined by the arcs of emission at 9.8-10.9~\kms~to the north and southeast of the center. At 8.9-9.4~\kms, an unrelated spur of emission appears to the north, and at 10-10.9~\kms, another unrelated spur is visible to the south. 

Nested inside of Shell~\nVbig, Shell~\nVsmall~is shown in the \co[12] channel maps in Figure~\ref{fig:shell_v380small_channels}. Shell~\nVsmall~is one of the most ideally symmetric shells in the catalogue, with a circular cavity that persists at higher velocities than the larger Shell~\nVbig. In fact, Shell~\nVsmall~includes some of the highest velocity CO emission in the southern half of Orion A. The ``smoke-ring'' structure of Shell~\nVsmall~is most clearly seen in the channel maps at 10.4-10.8~\kms.

\begin{figure*}
\centering
\includegraphics[width=10cm]{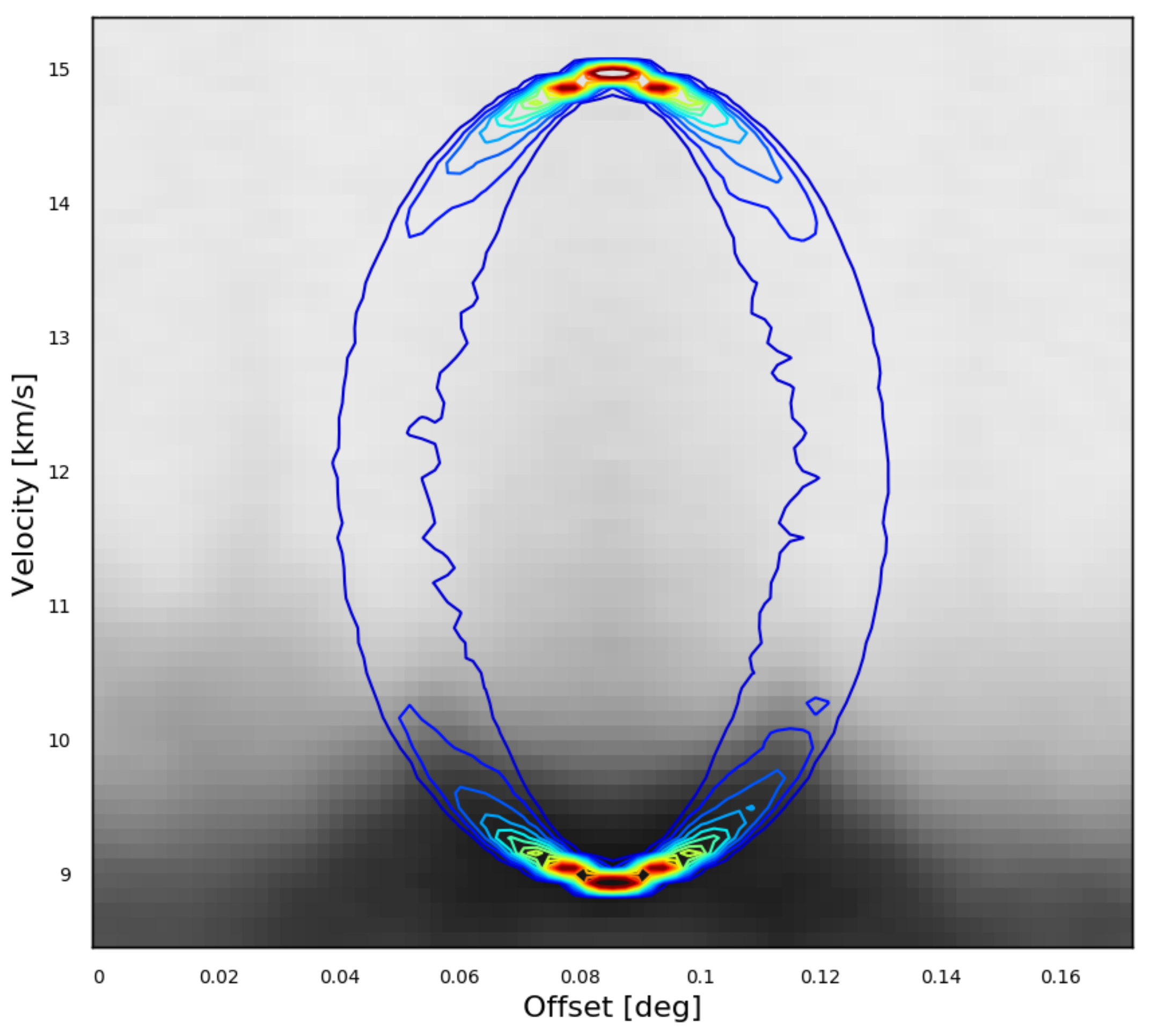}
\caption{Shell~\nVbig~\co[12] position-velocity diagram. We extract emission along 4 equally spaced slices through the center of the shell and average. Contours show the model that best represents the shell. The model parameters are given in Table~\ref{tab:shells}. Based on the U-shaped PV diagram, we only detect the near, approaching cap of the shell.}
\label{fig:shell_v380large_pv}
\end{figure*}

\begin{figure*}
\centering
\includegraphics[width=10cm]{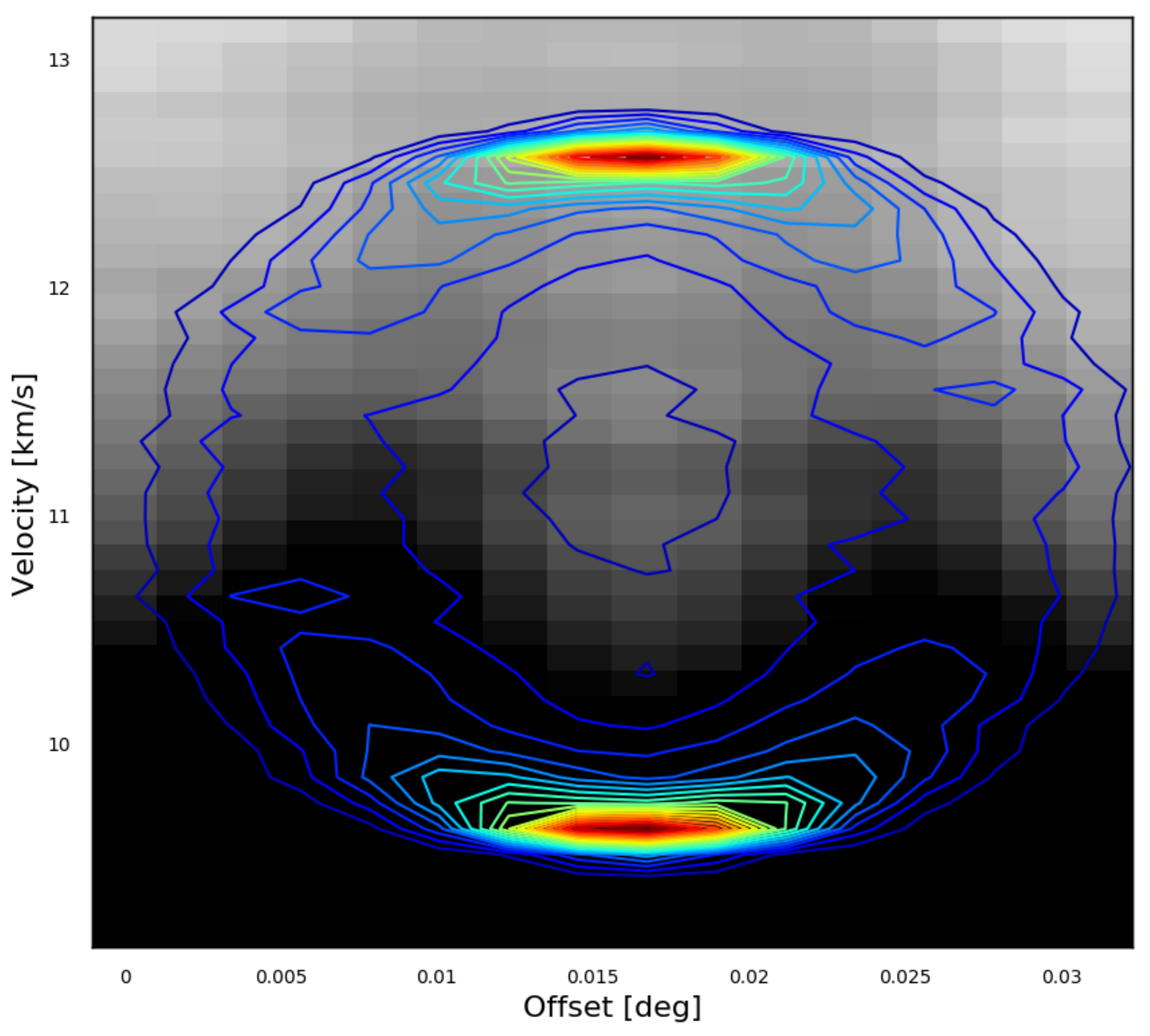}
\caption{Shell~\nVsmall~\co[12] position-velocity diagram. We extract emission along 4 equally spaced slices through the center of the shell and average. Contours show the model that best represents the shell. The model parameters are given in Table~\ref{tab:shells}.}
\label{fig:shell_v380small_pv}
\end{figure*}

\paragraph{Position-Velocity Diagram}
Figure~\ref{fig:shell_v380large_pv} and Figure~\ref{fig:shell_v380small_pv} show azimuthally averaged position-velocity diagrams of \co[12] towards Shell~\nVbig~and Shell~\nVsmall~respectively. We only detect the side of Shell~\nVbig~approaching us, lending its PV diagram a U-shaped morphology. Because we do not detect the shell through its entire velocity range, the expansion velocity is difficult to constrain. By contrast, Shell~\nVsmall~ is detected over most of its velocity range and shows a mostly complete ring structure in its PV diagram. The shell is very faint compared to the cloud emission at lower velocities, but its uniquely high central velocity separates it well from the rest of the cloud.

\begin{figure*}
\centering
\includegraphics[width=15cm]{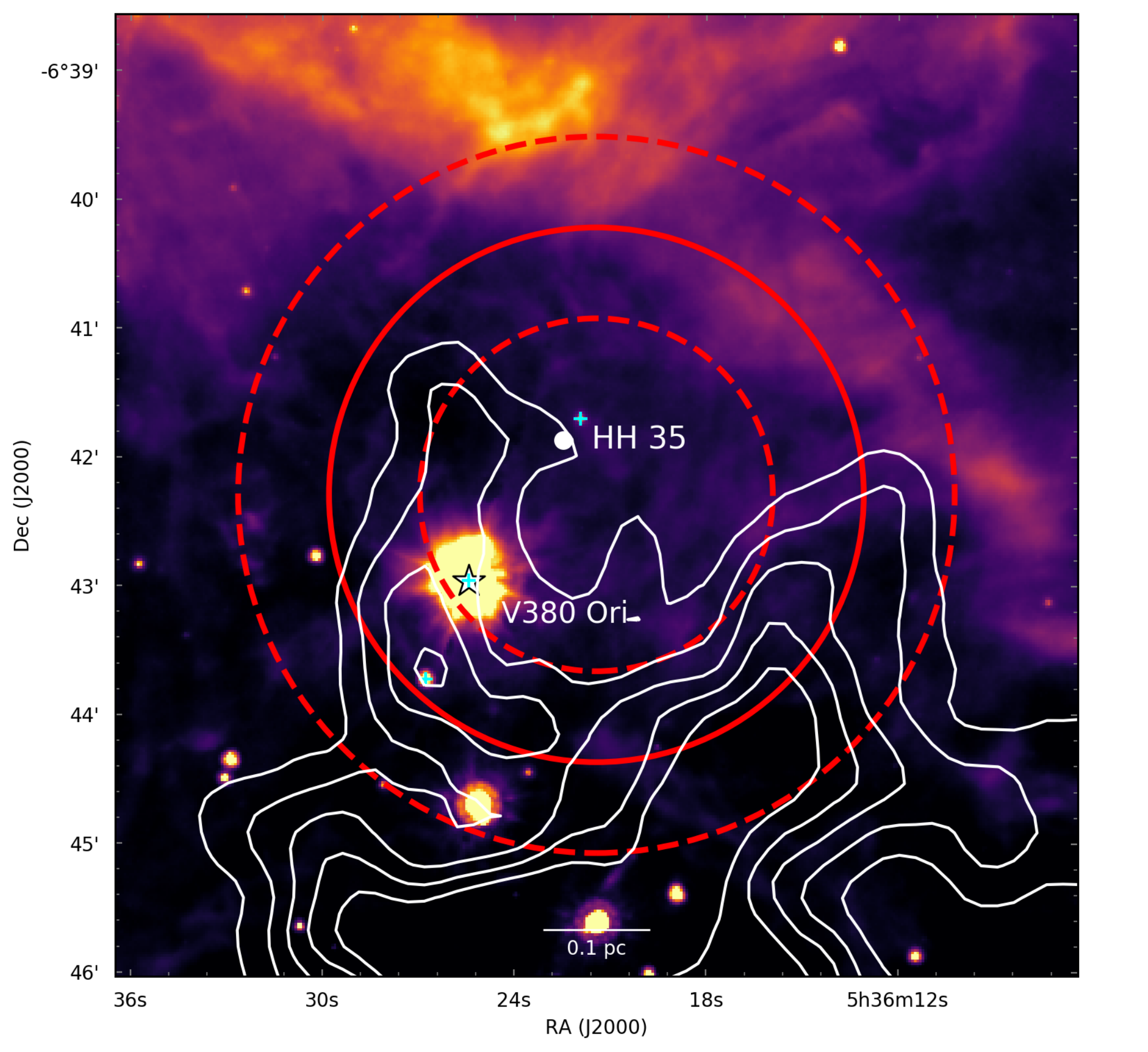}
\caption{Spitzer 8~$\mu$m map toward Shell~\nVbig. Contours show \co[13]~integrated from 9 to 11~\kms. Contours are drawn from 20 to 40$\sigma$ with steps of $5\sigma$, where $\sigma = 0.5$~K~\kms. The star indicates the Herbig B9e star V380 Ori. The cyan crosses indicate pre-main sequence stars from the Spitzer Orion catalog. The filled white circle indicates the Herbig-Haro object HH 35. The large solid circle and dashed annulus indicate the best-fit radius and thickness of the CO shell, respectively.}
\label{fig:shell_v380large_ir}
\end{figure*}

\paragraph{Infrared Nebulosity}
Figure~\ref{fig:shell_v380large_ir} shows 8~$\mu$m emission along with integrated \co[13] towards Shell~\nVbig. Much of the 8~$\mu$m emission in this area is concentrated to the north and west of the shell. This may be dust swept up by the part of the shell where CO is not seen or could be unrelated.

\begin{figure*}
\centering
\includegraphics[width=15cm]{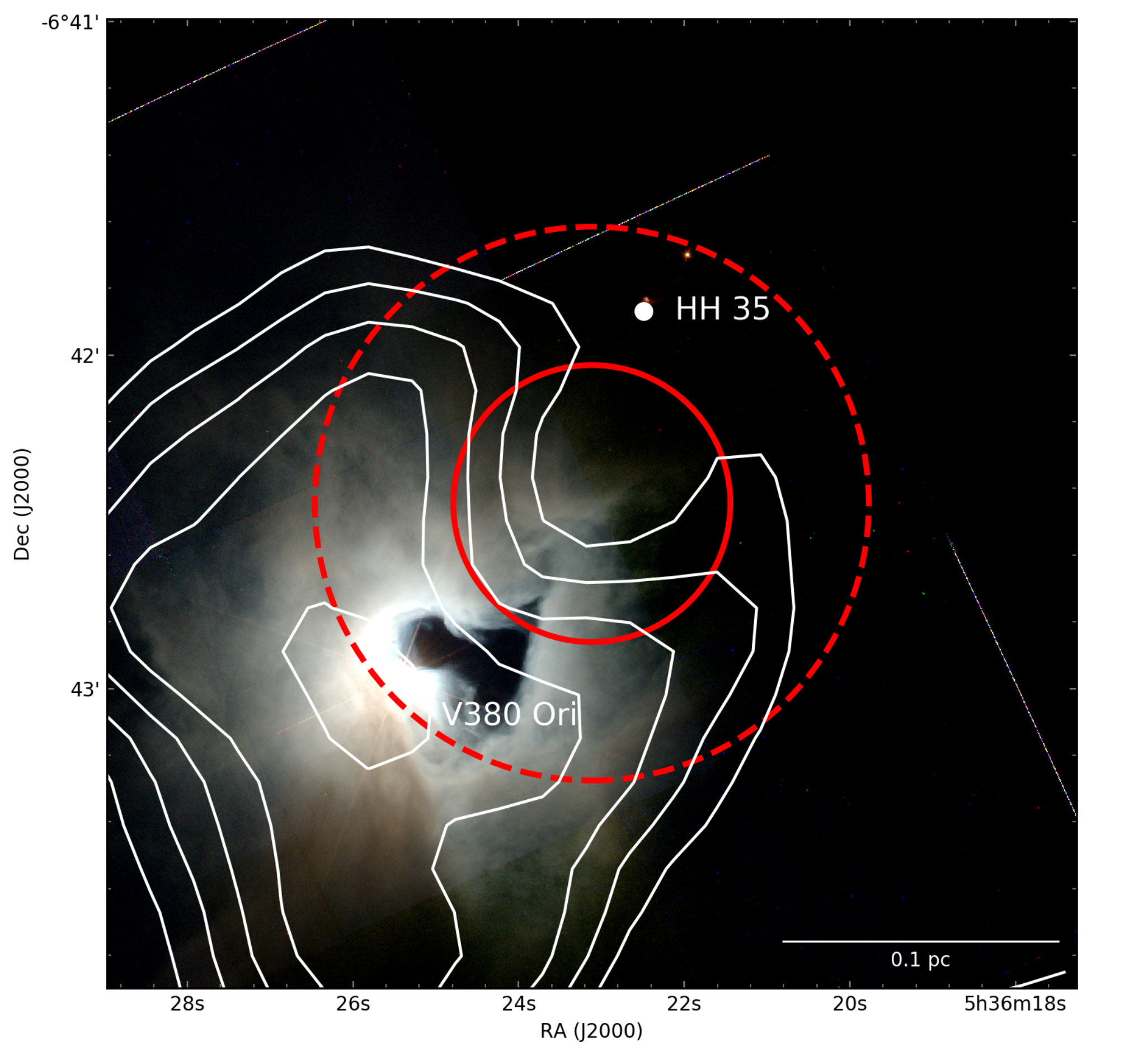}
\caption{\emph{HST WFC2} composite image toward Shell~\nVsmall~with F450W/F555W/F675W filters represented by red/green/blue colors respectively (Hubble Legacy Archive - Program 8548, PI: K. S. Noll). Contours show \co[12] integrated from $10.2$ to $12$~\kms. Contours are drawn from 20 to 32$\sigma$ with steps of $3\sigma$, where $\sigma = 0.7$~K~\kms. The bright nebulosity around V380 Ori is the reflection nebula NGC 1999. The dark feature is a cavity in the cloud, possibly excavated by outflows and/or winds from the Herbig B9e star V380 Ori. The Herbig-Haro object HH 35 (filled white circle) traces an outflow coming from the V380 Ori system which may help to shape the shell and cavity.}
\label{fig:shell_v380small_hst}
\end{figure*}

Figure~\ref{fig:shell_v380small_hst} shows integrated \co[12] toward Shell~\nVbig~with a three-color optical image taken from the Hubble Legacy Archive. There is no sign of related emission in the Spitzer images, but this shell is likely related to the dark cavity excavated by V380 Ori. We discuss this cavity in more detail below.

\paragraph{Potential Driving Sources}
The most likely driving source for both of these shells is the V380 Ori system. V380 Ori consists of a 1-3 Myr Herbig B9e star with a luminosity of 200 $L_\odot$ \citep{Rodriguez16}, an infrared companion identified by \cite{Leinert97}, a low-mass spectroscopic companion with a luminosity of 3~$L_\odot$ \citep{Alecian09}, and a fourth M5/6 companion \citep{Reipurth13}. 

V380 Ori is responsible for several Herbig-Haro flows, including the 5.3~pc long HH 222/1041 flow identified by \cite{Reipurth13}. This flow may have originated in a massive accretion event triggered by a dynamical decay of the quadruple stellar system. Based on the proper motion of HH 222, this event occurred less than 28,000 yr ago. The expansion time of a shell assuming uniform constant expansion is $t_{\rm{exp}} = R / v_{\rm{exp}}$. For Shell~\nVbig, $t_{\rm{exp}} \approx $80,000~yr. For Shell~\nVsmall, $t_{\rm{exp}} \approx $30,000~yr. If a shell's expansion has slowed over time it would be younger than this estimate. The same accretion-driven outburst that is responsible for the high-velocity large-scale Herbig-Haro flows may have caused an increased mass-loss rate and spherical wind that produced the expanding shells. The smaller-scale Herbig-Haro flows from V380 Ori are HH 1031/130 and HH 35, which may represent more recent dynamical interactions between the components of the V380 Ori system. Any of these interactions may have played a role in shaping the shells we see in this region.

\cite{Liu11} fit the SED of V380 Ori to derive a current infall rate from the envelope of $2\times10^{-6}$~\Mdot~ and a disk accretion rate of $3\times10^{-9}$~\Mdot. Typically, the mass-loss rate of a protostar is expected to be about 10-30\% of the accretion rate \citep[e.g.,][]{Pudritz07,Mohanty08}. This implies a mass-loss rate of $3\times10^{-10}$~to~$9\times10^{-10}$~\Mdot. Shell~\nVbig~requires a wind mass-loss rate of a few $10^{-7}$~\Mdot~and Shell~\nVsmall~requires $10^{-8}$~to~$10^{-7}$~\Mdot~(see Table~\ref{tab:physics}). An accretion-driven outburst like the one discussed above could strengthen the wind enough to produce the expanding shell \citep[see][\S~4.3.2]{Offner15}. Such wind enhancements over short timescales ($\approx 0.1$~Myr) could have powered the shells despite the much lower current mass-loss rate. We discuss this mechanism more in Section~\ref{sec:discussion}.

Adjacent to the NGC 1999 reflection nebula is a dark cavity in the cloud indicated by a deficit in far-infrared emission coupled with lower extinctions of background stars through this part of the nebula \citep{Stanke10}. Figure~\ref{fig:shell_v380small_hst} shows that the CO shell is offset from the optical cavity by about 0.1~pc. \citet{Stanke10} speculate that that the outflow driving HH 35 and H$_{2}$~2.12$\mu$m shock SMZ 6-8 (\cite{Stanke02}) is responsible for carving out the northern part of this cavity. Near this dark cavity, \citet{Corcoran95} found a cavity in H$\alpha$ that may also be related to the V380 outflows. In this scenario, the shell may be considered an evolved state of the wide-angle outflow cavities observed around outbursting pre-main sequence stars \citep{Ruiz-Rodriguez17,Principe18}.

\subsection{A Shell Centered on LP Ori}\label{sec:shell_LPOri}
Shell~\nLP~is about 0.1~degrees~($0.7$~pc) southwest of OMC~1. At its center lies the pre-main sequence B2 star LP Ori.

\begin{figure*}
\centering
\includegraphics[width=13cm]{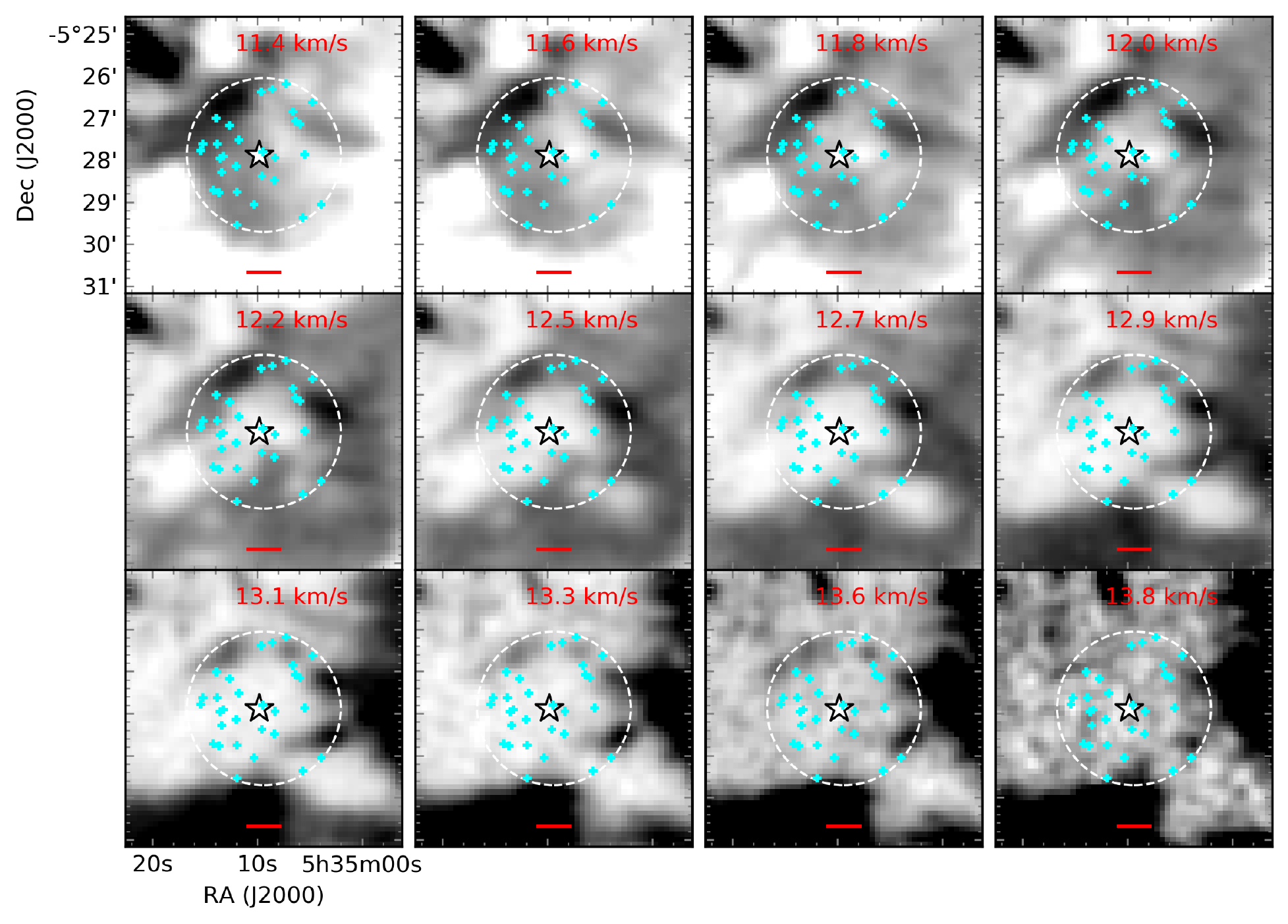}
\caption{\co[12] channel maps of Shell~\nLP, associated with LP~Ori (white star). Symbols are the same as Figure~\ref{fig:shell_TOri_channels}. The best-fit radius is shown as a dashed white circle. The red scalebar has a length of 0.1~pc.}
\label{fig:shell_LPOri_channels}
\end{figure*}

\paragraph{CO Channel Maps}
Figure~\ref{fig:shell_LPOri_channels} shows \co[12] channel maps toward Shell~\nLP. The distinctly circular shell is highlighted by emission along the rim to the northeast and west. The bright unrelated emission in the northeast corner of the channel maps is associated with the Orion Bar photodissociation region (PDR). Shell~\nLP~is superimposed upon a larger CO expansion seen to the south and west at 13 to 13.5~\kms. First identified by \citet{Loren79} and \citet{Heyer92}, this $\approx~2$~pc CO shell traces the southern edge of the Orion Nebula HII region and is likely being driven by the expansion of the HII region into the molecular cloud behind it. The HII-driven CO expansion can also be seen in the vicinity of the T~Ori shell (Section~\ref{sec:shell_TOri}).

\begin{figure*}
\centering
\includegraphics[width=10cm]{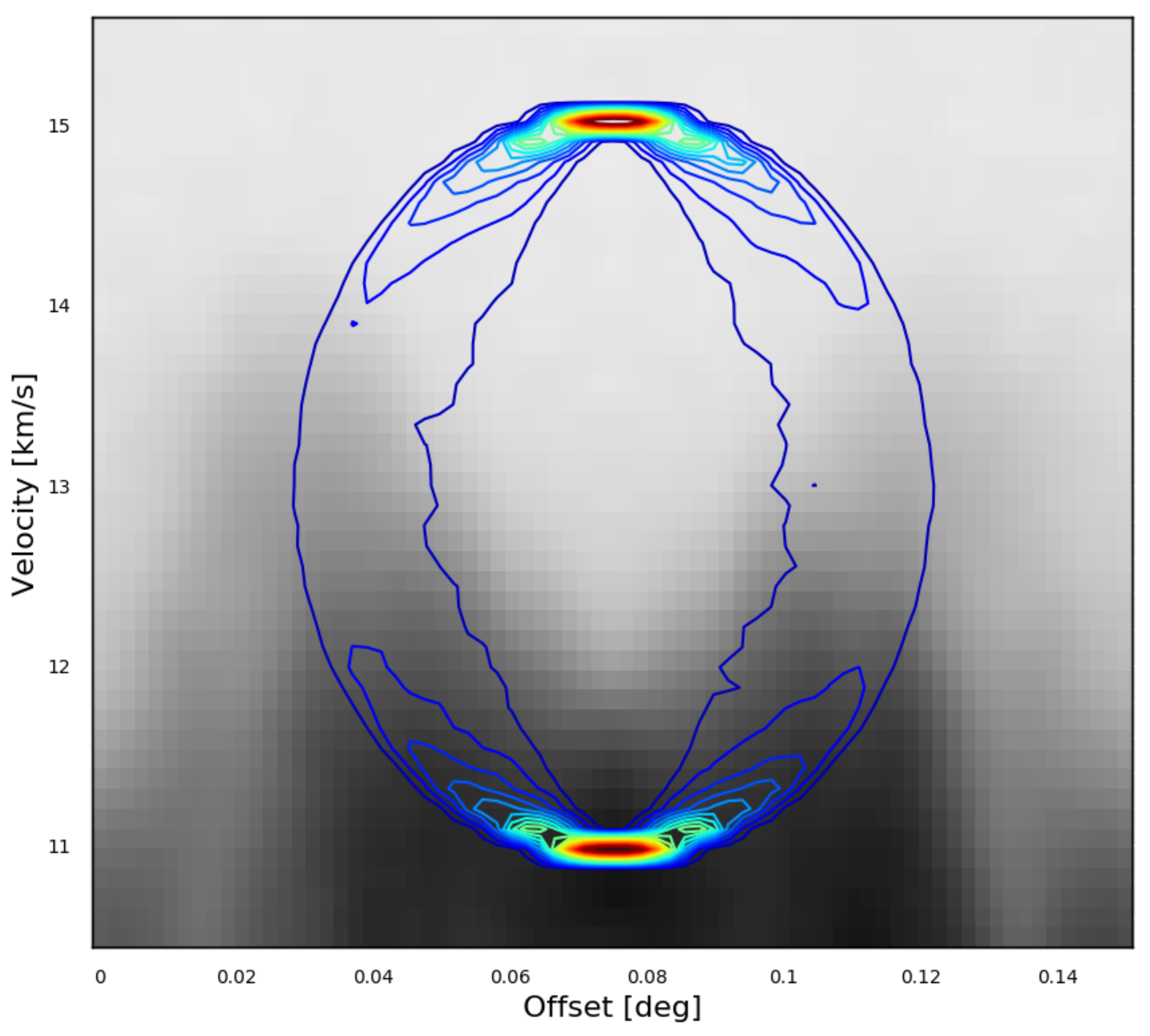}
\caption{Azimuthally averaged position-velocity diagram of \co[12] emission toward Shell~\nLP. We extract emission along 4 equally spaced slices through the center of the shell and then average. Contours show the model that best represents the shell. The model parameters are given in Table~\ref{tab:shells}. Based on the U-shaped PV diagram, we only detect the near, approaching cap of the shell.}
\label{fig:shell_LPOri_pv}
\end{figure*}

\paragraph{Position-Velocity Diagram}
Figure~\ref{fig:shell_LPOri_pv} shows the azimuthally-averaged position-velocity diagram of \co[12] toward Shell~\nLP. The U-shape of the PV diagram indicates that the expanding shell is only detected at velocities blueward of the central shell velocity. Thus, we only see the emission on the near side of the shell while the far side of the shell has broken out of the cloud.

\begin{figure*}
\centering
\includegraphics[width=15cm]{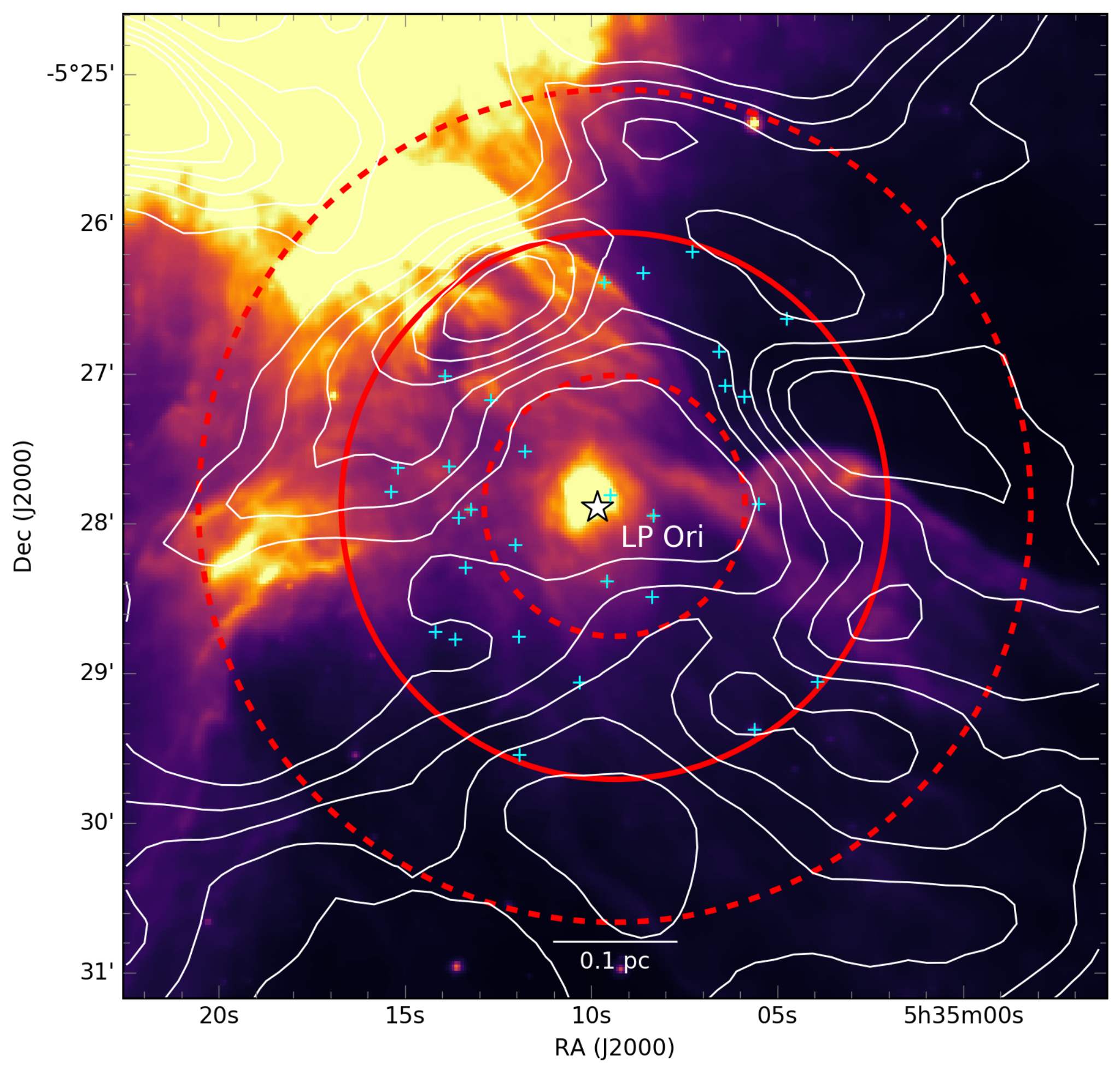}
\caption{Spitzer 8~$\mu$m map toward Shell~\nLP. Contours show \co[12]~integrated from 11.7 to 13.4~\kms. Contours are drawn from 35 to 75$\sigma$ with steps of $8\sigma$, where $\sigma = 0.7$~K~\kms. The white star indicates the likely driving source - pre-main sequence B2V star LP Ori. Cyan crosses indicate pre-main sequence stars from the Spitzer Orion catalog. The large solid circle and dashed annulus indicate the best-fit radius and thickness of the CO shell, respectively.}
\label{fig:shell_LPOri_ir}
\end{figure*}

\paragraph{Infrared Nebulosity}
Figure~\ref{fig:shell_LPOri_ir} shows 8~$\mu$m emission along with integrated \co[12] toward Shell~\nLP. The shell is located near the bright infrared emission from the Orion Nebula in the northeastern corner of Figure~\ref{fig:shell_LPOri_ir}). This complicates any analysis of dust emission correlated to the CO shell, but infrared nebulosity along the north and east of the shell rim may trace dust swept up by the shell. The central star LP Ori is shrouded in dust emission, a sign that it is still associated with its birth cloud.

\paragraph{Potential Driving Sources}
Located at the center of Shell~\nLP, LP Ori (HD 36982) is a B2V pre-main sequence star \citep{Hillenbrand13}. While it lacks spectral emission lines, LP Ori was classified as a Herbig Be star by \cite{Manoj02} on the basis of its infrared excess. LP Ori is one of the $\sim7\%$ of Be stars with an organized magnetic field \citep{Alecian17}, as measured by polarimetry \citep{Petit08,Alecian13}. 

Using model stellar evolutionary tracks, \cite{Alecian13} report a mass of $7~M_\odot$ and age of 0.2~Myr. The age of LP Ori is consistent with Shell~\nLP's expansion time of 0.1 Myr.

Using the mass-loss recipe of \cite{Vink00}, \cite{Naze14} estimates LP Ori's mass-loss rate at $10^{-9}$~\Mdot, or 2-3 orders of magnitude lower than the necessary wind mass-loss rate needed to drive the observed shell (Table~\ref{tab:physics}). In order to produce the required momentum, LP Ori may have undergone a burst of accretion-driven mass-loss.

\section{Impact of Shells on Cloud}\label{sec:impact}

\subsection{Measuring Mass, Momentum, and Kinetic Energy}\label{sec:impact_method}

We measure the mass, momentum, and kinetic energy of the shell candidates following methods laid out in \cite{Arce01}, \cite{Arce10}, and \cite{Arce11}. We briefly describe our method here. For more details, see \cite{Dunham14a} and \cite{Zhang16a}.

To extract the shell from the spectral cube, we first construct a mask using a model cube generated from the best fit parameters as described in Section~\ref{sec:model}. We extract each shell multiple times using sets of model parameters spanning the ranges given in Table~\ref{tab:shells} to estimate the uncertainty on the derived physical properties. 

Where \co[13] is detected at~$5\sigma$ we assume it is optically thin and use it in the mass calculation. Where \co[13] is not significant but \co[12] is detected at~$5\sigma$, we compute an opacity correction to \co[12] using the \co[12]/\co[13] ratio in the vicinity of the shell. This correction is detailed below.

Assuming that \co[12] and \co[13] are both in LTE, have the same excitation temperature, and \co[13] is optically thin, the ratio between the \co[12] and \co[13] brightness temperature is 
\begin{equation}\label{eq:Tratio}
\frac{T_{\rm \co[12]}}{T_{\rm \co[13]}} = \frac{[\rm{\co[12]}]}{[\rm{\co[13]}]} \frac{1 - e^{-\tau_{12}}}{\tau_{12}}.
\end{equation}
[\co[12]]/[\co[13]] is the abundance ratio, assumed to be 62~\citep{Langer93}, and $\tau_{12}$ is the opacity of \co[12]. We measure the velocity-dependent ratio between the \co[12] and \co[13] brightness temperature, averaging over an area around each shell. Using this ratio and Equation~\ref{eq:Tratio} we compute the opacity correction factor $\tau_{12} / (1 - e ^{-\tau_{12}})$ at each velocity channel for each shell. We multiply the observed $T_{\rm \co[12]}$ in each shell voxel by this factor to correct for opacity.

We add the shell voxels with $5 \sigma$ \co[13] to the shell voxels without $5 \sigma$ \co[13] but having $5 \sigma$ \co[12]. Integrating each, we compute the column density of $\rm H_2$ using equation A6 in \citet{Zhang16a}:

\begin{equation}\label{eq:dNdv}
\frac{dN}{dv} = \left(\frac{8\pi k \nu^2_{ul}}{h c^3 A_{ul} g_u}\right) Q_{\rm rot}(T_{\rm ex})~ e^{E_u / kT_{\rm ex}} \frac{T_R(v)}{f}
\end{equation}

where $\nu_{ul}$ is the frequency of the transition, $A_{ul}$ is the Einstein A coefficient, $E_u$ is the energy of the upper level, $g_u$ is the degeneracy of the upper level, $Q_{\rm rot}$ is the partition function (calculated to $j=100$), $T_{\rm ex}$ is the excitation temperature, $T_R(v)$ is the brightness temperature of the CO line (opacity-corrected \co[12] or \co[13]), and $f$ is the abundance ratio of $\rm H_2/\rm CO$. For \co[12], $f = 1 \times 10 ^{-4}$. For \co[13], $f = 1 \times 10 ^{-4} / 62$ \citep{Frerking82}.

An excitation temperature is calculated for each shell by assuming \co[12] is optically thick. We estimate $T_{\rm ex}$ with the peak brightness temperature of the average \co[12] spectrum in the vicinity of each shell, using the equation from \cite{Rohlfs96}:

\begin{equation}\label{eq:Tex}
T_{\rm ex} = \frac{5.53}{ln(1 + [5.53/(T_{\rm peak} + 0.82)])}
\end{equation}

The mass of molecular hydrogen in each voxel is then $M_{\rm H_2} = m_{\rm H} \mu_{\rm H_2} A_{\rm pixel} N_{\rm H_2}$ where $m_{\rm H}$ is the mass of a hydrogen atom, $\mu_{\rm H_2}$ is the mean molecular weight per hydrogen molecule, and $A_{\rm pixel}$ is the spatial area subtended by each pixel at the distance of the cloud \citep[$414$~pc for Orion A,][]{Menten07}.

We find the total mass of a shell by adding the mass in every shell voxel. We use this mass to calculate the momentum $P = M_{\rm H_2}v_{\rm exp}$ and kinetic energy $E = 0.5 M_{\rm H_2}v_{\rm exp}^2$, assuming that the shell is expanding uniformly at the model's expansion velocity.

For each shell, we report best-fit values as well as lower and upper limits on mass, momentum, and kinetic energy in Table~\ref{tab:physics}. The lower limits are found by using the lower limits on the model radius, thickness, and expansion velocity reported in Table~\ref{tab:shells} to extract the voxels in the shell. We compute multiple models with these lower limits at central velocities ($v_0$) spanning the range reported in Table~\ref{tab:shells}. The median of this set of models is the lower limit reported in Table~\ref{tab:physics}. We compute the best-fit values and upper limits in the same way except with the best-fit values and upper limits on model radius, thickness, and expansion velocity from Table~\ref{tab:shells}.

Unrelated emission overlaps the shells in many of the channel maps. This may contaminate the derived masses of the extracted shells. We use models instead of extracting shell voxels by hand, accepting some contamination from extraneous cloud emission in order to report consistent and reproducible masses. As described in Section~\ref{sec:model}, the uncertainties on the model parameters (Table~\ref{tab:shells}) are large and reflect the most extreme models that resemble the observed shells, so any contamination in shell mass (and all values derived from shell mass) should fall within the uncertainties reported in Table~\ref{tab:physics}.

\subsection{Mass, Momentum, and Kinetic Energy Statistics}\label{sec:impact_stats}
\begin{figure}
\centering
  \includegraphics[width=1.1\columnwidth]{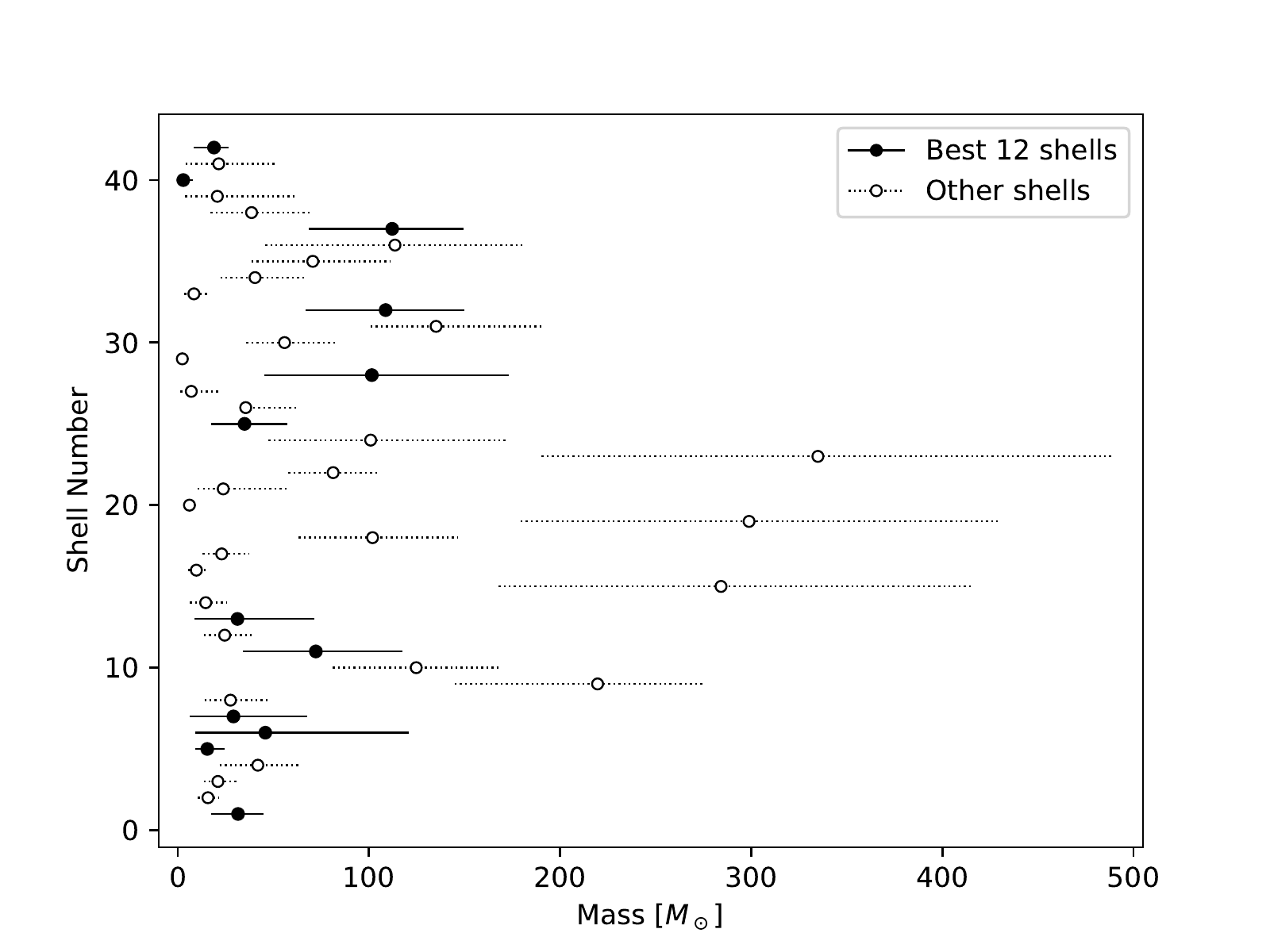}
\caption{Range in mass of each shell candidate. The points represent masses derived from best-fit model parameters. Upper and lower limits are calculated using the upper and lower limits on the model parameters (see Section~\ref{sec:impact_stats}). The 12 most robust shells are shown as filled circles.}
\label{fig:mass_range}
\end{figure}

\begin{figure}
\centering
\includegraphics[width=1.1\columnwidth]{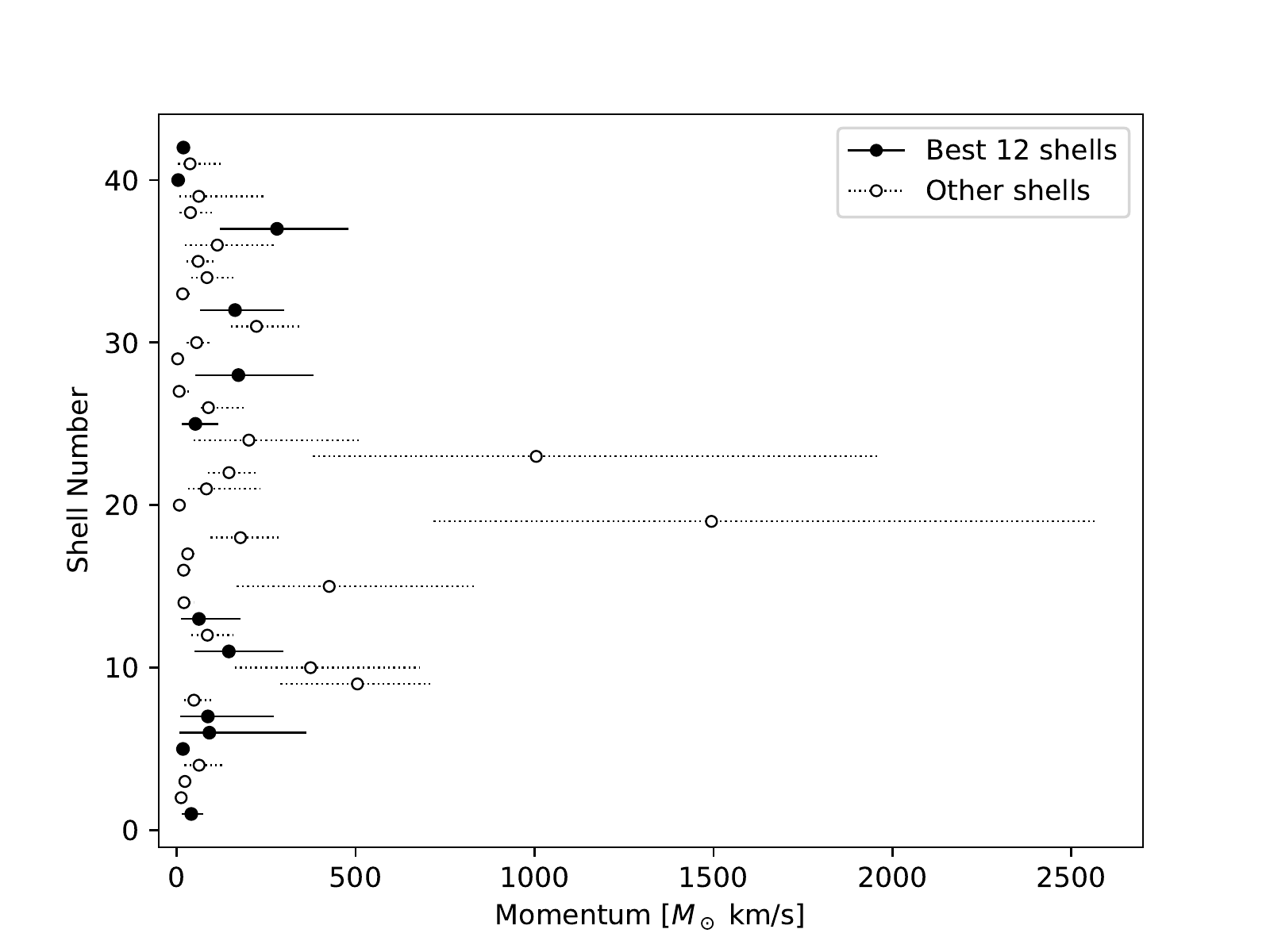}
\caption{Same as Figure~\ref{fig:mass_range} except plotting the range in momentum of each shell.}
\label{fig:momentum_range}
\end{figure}

\begin{figure}
\centering
\includegraphics[width=1.1\columnwidth]{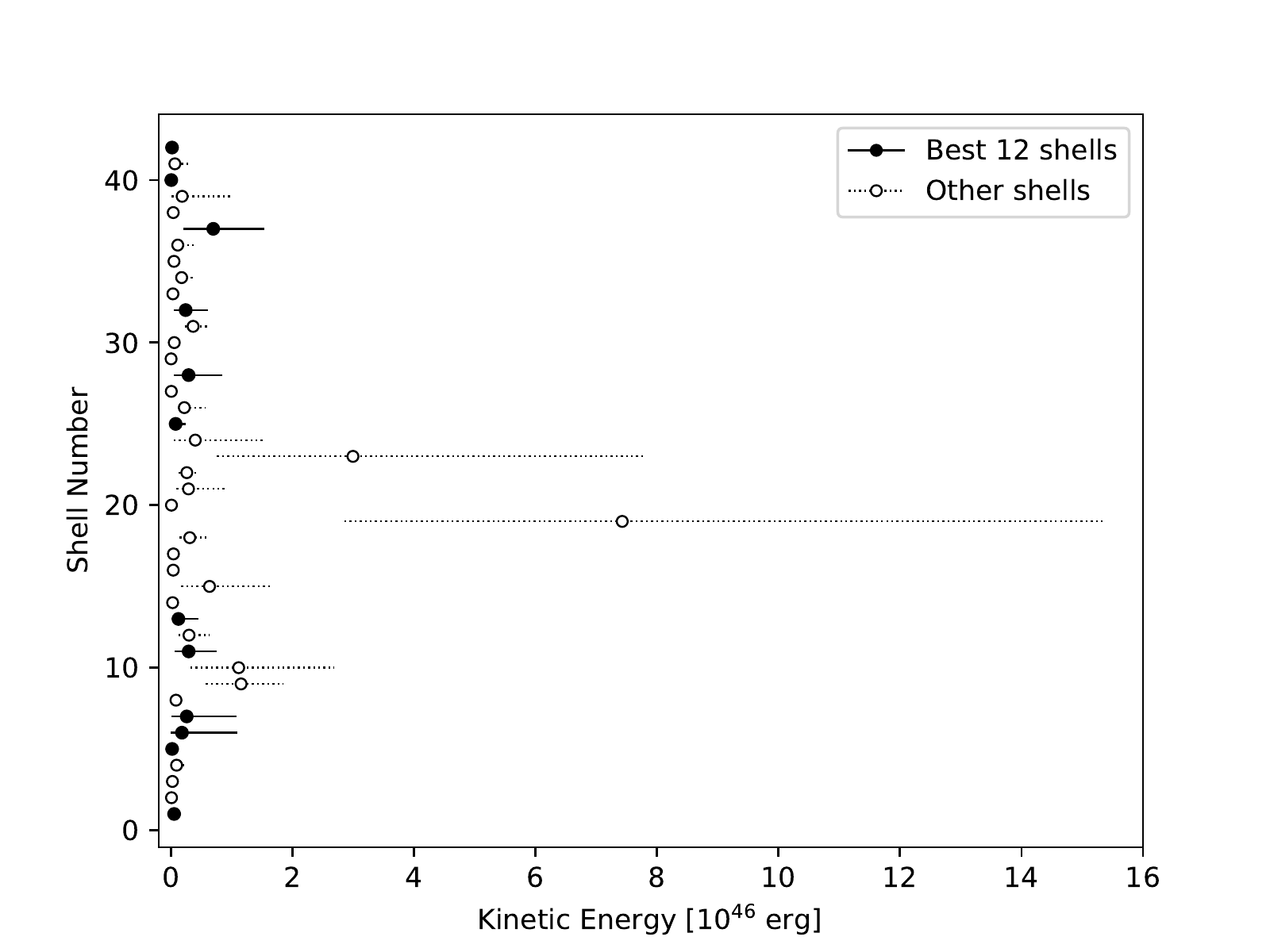}
\caption{Same as Figure~\ref{fig:mass_range} except plotting the range in kinetic energy of each shell.}
\label{fig:energy_range}
\end{figure}

Figures~\ref{fig:mass_range},~\ref{fig:momentum_range},~and~\ref{fig:energy_range} show the distribution of mass, momentum, and energy for the full shell sample and the 12 most robust shells which meet all criteria in Table~\ref{tab:criteria}. We show the range of physical parameters derived for each shell using the lower limit, best-fit, and upper limits on the model parameters. 

Table~\ref{tab:impact} reports the total kinetic energy of the shells in Orion A. The total mass, momentum, and energy contained within shells in Orion A are similar to the cumulative totals of the Perseus cloud shells reported by~\citet{Arce11}.

\section{Discussion}\label{sec:discussion}
\subsection{The Impact of Shells on the GMC}\label{sec:comparison}

To compare the impact of the shells on the cloud with protostellar outflows and cloud turbulence, we split Orion A into several subregions, shown in Figure~\ref{fig:12co_peak_shells}. The North subregion covers the OMC 2 and OMC 3 areas of the molecular cloud, as well as the southern portion of the HII region NGC 1977 \citep{Peterson08,Davis09}. The Central subregion includes the Orion Bar \citep{Goicoechea16}, Orion KL and the OMC 1 explosive outflow \citep{Bally17}. The South subregion covers OMC 4 and OMC 5 \citep{Johnstone06,Buckle12,Davis09}. The L1641N subregion covers the L1641 North cluster and the reflection nebula NGC 1999 powered by Herbig Be9 star V380 \citep{Davis09,Nakamura12}. Shells are assigned to the subregion containing their center.

\subsubsection{Comparing Shells and Protostellar Outflows}\label{sec:outflows}

To assess the relative importance of feedback mechanisms, we compare the kinetic energy and momentum injected by the shells and by protostellar outflows gathered from the literature. Table~\ref{tab:impact} summarizes the impact of protostellar outflows in Orion A. We detail below the outflows considered in each subregion.

\paragraph{North}\label{sec:outflows_north}
In the North, outflows in OMC~2 and OMC~3 were observed by \citet{Williams03} and later expanded by \citet{Takahashi08}. We estimate the kinetic energy and mechanical luminosity of these outflows using the velocities, masses, and dynamical times reported in \citet{Takahashi08}~Table~3. The 15 outflows in OMC~2/3 contain a total kinetic energy of $6.8 \times 10^{45}$~erg, mechanical luminosity of $2 \times 10^{34}$~erg~s$^{-1}$, and momentum injection rate of $2\times10^{-3}~M_\odot$~\kms~yr$^{-1}$.

\paragraph{Central}\label{sec:outflows_central}
In the Central region, the OMC~1 explosive outflow dominates. \cite{Bally17} made detailed measurements of the outflowing gas using ALMA. The energy [momentum] of this outflow has been estimated at $4 \times 10^{46}$~erg [$160~ M_\odot$~\kms] \citep{Snell84} to $4 \times 10^{47} $~erg [$1257~M_\odot$~\kms] \citep{Kwan76}. We adopt an average of $10^{47}$~erg [$730~M_\odot$~\kms]. \cite{Snell84} found a dynamical time of $1500$~yr, corresponding to a mechanical luminosity of $2.1 \times 10^{36}$~erg~s$^{-1}$ and a momentum injection rate of $0.47~M_\odot$~\kms~yr$^{-1}$.  

About 100$\arcsec$~south, another high-velocity outflow was identified by \cite{Rodriguez-Franco99} in the OMC1-South region. \cite{Zapata05} measured a total energy of $4.6 \times 10^{46}$~erg, mechanical luminosity of $2.3 \times 10^{36}$~erg~s$^{-1}$, and momentum injection rate of $9.3 \times 10^{-2}$~\Msun~\kms~yr$^{-1}$. These two high velocity outflows dominate the Central subregion.

\paragraph{South}\label{sec:outflows_south}
In the South, we could not find any systematic study of outflows. As part of the Gould's Belt survey, \cite{Buckle12} identify three outflows in their \co[13] map of the OMC4 region, but do not measure the energetics of these outflows. \cite{Berne14} include the OMC1-South outflow discussed above in their assessment of the energetics of this region, but it is clearly contained in our Central subregion.

\paragraph{L1641N}\label{sec:outflows_l1641n}
In L1641-N, \cite{Stanke07} detected a sample of outflows which was expanded on by \cite{Nakamura12}. \cite{Nakamura12} measure five outflows in L1641N with a total mass, momentum, and energy of 13~\Msun, 80~\Msun~\kms, and $5.46 \times 10^{45}$~erg. Assuming an outflow dynamical time of a few $\times~10^{4}$~yr \citep{Nakamura12}, the mechanical luminosity of these outflows is $10^{34}$~erg~s$^{-1}$ and the momentum injection rate is $2 \times 10^{-3}$~\Msun~\kms~yr$^{-1}$.

\cite{Morgan91} measured three other outflows south of the L1641N cluster but within our L1641N subregion. Two of these outflows are apparently associated with the Herbig Haro objects HH 1/2 and V380. \citet{Morgan91} calculate upper and lower limits on the energy of these outflows. The lower limit only accounts for emission in the high-velocity wings of the outflow spectrum. The upper limit attempts to account for the outflow emission at very low velocities (presumably buried under the line core) by assuming that the molecular outflow emission at each velocity channel in the line core is equal to the emission of the lowest velocity channel in the wing. For these three outflows, we adopt an average of the lower and upper limits for a total energy of $7.4 \times 10^{45}$~erg, mechanical luminosity of $6.9 \times 10^{33}$~erg~s$^{-1}$, and momentum injection rate of $2.6 \times 10^{-3}$~\Msun~\kms~yr$^{-1}$.

\subsubsection{Cloud Kinetic Energy}\label{sec:cloudenergy}
We measure the kinetic energy in the molecular cloud in each of the subregions described above. We follow the method in Section~\ref{sec:impact_method} to calculate the H$_2$ mass in each pixel using \co[13] (when present) and opacity-corrected \co[12]. We estimate the velocity dispersion of the molecular gas following the method of \cite{Li15}. The one-dimensional velocity dispersion $\sigma_{\rm los}$ is computed from the second-moment map of \co[13]. The three-dimensional velocity dispersion $\sigma_{\rm 3D}$, assuming an isotropic turbulent velocity field, is equal to $\sqrt{3} \sigma_{\rm los}$. The kinetic energy in each pixel is $(1/2) \rm{M}_{H_2} \sigma_{\rm 3D}^2$. The total kinetic energy of a subregion is a sum of the kinetic energy in each of the subregion's pixels. Table~\ref{tab:impact} compares the kinetic energy of the cloud to the energy injected by the shells in each subregion.

\subsubsection{Shells and Turbulence}\label{sec:turb}
\paragraph{Energy Injection and Dissipation}\label{sec:turb_energy}
In the previous section we showed that the total energy contained within expanding shells is a significant fraction of the turbulent energy in the Orion A cloud. But in order to maintain this turbulence, the shells must provide this energy at a rate greater than or equal to the turbulent energy dissipation rate.

The turbulent energy dissipation rate $\dot E_{\rm turb}$ is given by the total turbulent energy $E_{\rm turb} = 5.8 \times 10^{47} \rm{erg}$ divided by the dissipation timescale $t_{\rm diss}$. \citet{Arce11} estimates $t_{\rm diss} = 5$~Myr in Perseus using the method of \cite{Mac-Low99}.

Alternatively, \citet{McKee07} show that the dissipation time of a homogeneous isotropic turbulent cloud with diameter $d$ and one-dimensional velocity dispersion $\sigma_{\rm los}$ is:
\begin{equation}\label{eq:tdiss}
t_{\rm diss} = 0.5 \frac{d}{\sigma_{\rm los}}
\end{equation}
We use the geometric average of the cloud length and width in the plane of the sky to estimate $d \approx 12$~pc. The median $\sigma_{\rm los}$ of \co[13] is 1.7~\kms. Using Equation~\ref{eq:tdiss}, we estimate $t_{\rm diss} \approx 3.5$~Myr in Orion A. With these assumptions, the turbulent energy dissipation rate is $10^{34}~\rm{erg}~\rm{s}^{-1}$, a factor of a few higher than that found in the Perseus \citep{Arce11} and Taurus \citep{Li15} molecular clouds. We repeat this procedure in each subregion, estimating $d = (4, 2, 5, 4)$~pc and $\sigma_{\rm los} = (1.6, 1.7, 1.6, 1.6)$~\kms~in the (North, Central, South, L1641N) subregions respectively.

The mechanical luminosity of a shell $\dot{E}_{\rm shell}$ can be simply estimated by dividing the shell energy by the expansion time of the shell $t_{\rm exp}$. Assuming the shell has expanded at a constant rate, $t_{\rm exp} = R~/~v_{\rm exp}$. For the purposes of this calculation, we use the best-fit radius and expansion velocity for each shell reported in Table~\ref{tab:shells}. The mechanical luminosity of each shell is reported in Table~\ref{tab:physics} and the total mechanical luminosity of the shells is reported in Table~\ref{tab:impact}.

In the North, the mechanical luminosity of shells is twice the turbulent dissipation rate and a factor of five lower than the outflow injection rate. In the Central subregion, the shells contain 70\% the power of turbulent dissipation and contribute a small fraction of the outflow injection rate which is dominated by the Orion KL explosive outflow. The shells have the most impact in the South, where the total shell luminosity is a factor of nine higher than the turbulent dissipation rate (we found no outflows in the South).\footnote{The South is dominated by two outliers: Shell 19 and 23. These are two of the largest shells in the catalog, with high expansion velocities. The physical quantities for these shells are likely to be more contaminated by unrelated emission compared to the other shells. Removing the contribution from Shell 19 and 23 reduces the shell luminosity in the South to about 30\% higher than the turbulent energy dissipation rate. See Table~\ref{tab:impact} for more details.} In L1641N, the shell luminosity is comparable to the turbulent dissipation rate and a factor of a few lower than the outflow injection rate. The shells contain enough power to counteract the turbulent dissipation rate in all but perhaps the Central subregion. A similar result was found for the shells in Perseus by~\citet{Arce11}. In Taurus, \citet{Li15} found that shells inject energy at about 2-10$\times$~the turbulent dissipation rate. 

\paragraph{Momentum Injection and Dissipation}\label{sec:turb_momentum}\

Because shells and outflows are momentum-driven, \citet{Nakamura14} compare the outflow momentum injection rate to the momentum dissipation rate in several clouds. We find the momentum dissipation rate of the cloud regions using Equation 4 in \citet{Nakamura14}:

\begin{eqnarray}\label{eq:dPdt}
\frac{dP_{\rm turb}}{dt} = 6.4 \times 10^{-4}~M_\odot~\rm{km}~\rm{s}^{-1}~\rm{yr}^{-1}\times \\
\left(\frac{M_{\rm cl}}{500~M_\odot}\right) \times \left(\frac{R_{\rm cl}}{0.5~\rm{pc}}\right)^{-1} \times \left(\frac{\sigma_{\rm los}}{\rm{km}~\rm{s}^{-1}}\right)^2 \nonumber
\end{eqnarray}

$M_{\rm cl}$ is the mass of the cloud subregion, $R_{\rm cl}$ is the radius of the cloud subregion, and $\sigma_{\rm los}$ is the line-of-sight velocity dispersion of the cloud subregion. We use the same estimates as the energy dissipation calculation, $R_{\rm cl} = (2,1,2.5,2)$~pc and $\sigma_{\rm los} = (1.6, 1.7, 1.6, 1.6)$~\kms~, and find $M_{\rm cl} = (4048, 3736, 5001, 5196)~$\Msun~for the North, Central, South, and L1641N subregion, respectively. Because the method of \citet{Nakamura14} is intended for the clump scale, we do not apply Equation~\ref{eq:dPdt} to the entire cloud, but only report the momentum dissipation rates of the subregions in Table~\ref{tab:impact}.

We compare the momentum dissipation rate of the cloud subregions to the momentum injection rates of the outflows (reported in Section~\ref{sec:outflows}) and shells. As with the mechanical luminosity, we calculate a shell's momentum injection rate by dividing the shell momentum by its expansion time. The momentum injection rate of each shell is reported in Table~\ref{tab:physics} and the total shell momentum injection rate is reported in Table~\ref{tab:impact}.

In the North, the shells inject momentum at about three times the rate of outflows and twice the dissipation rate. In the Central subregion, shells inject enough momentum to counteract dissipation but are again dominated by the massive outflows in Orion KL. In the South, shells inject momentum at seven times the dissipation rate.\footnote{If the two outliers Shell 19 and 23 are removed, the total shell momentum injection rate in the South is a factor of two higher than the turbulent dissipation rate. See note $d$ in Table~\ref{tab:impact} for more details.} In L1641N, the shell momentum injection rate is twice that of the outflows and twice the  dissipation rate. 

The shells inject more momentum into the cloud than outflows except in the Central subregion, which is dominated by high velocity outflows. The momentum injection by shells and outflows is greater than the momentum dissipation rate throughout the cloud and can thus maintain the cloud turbulence.

\subsection{Shell Driving Mechanisms}\label{sec:driving}
What powers the shells? \cite{Arce11} and \cite{Offner15} consider protostellar outflows, turbulent voids, and stellar winds. Protostellar outflow cavities are generally collimated but could appear circular if viewed on-axis. Because most outflows are highly collimated, the momentum on the plane of the sky is a small fraction of the total outflow momentum. \cite{Offner15} estimate the outflow rates required to drive a typical shell would be several orders of magnitude higher than observed. Wide-angle outflows are sometimes observed around pre-main sequence stars \citep[e.g.][]{Ruiz-Rodriguez17,Principe18}. Such an outflow would not need to be viewed on-axis and may help explain structures like Shell~\nVsmall~(Section~\ref{sec:shell_v380}).

Random turbulent voids may masquerade as feedback-driven shells. \citet{Offner15} find that CO voids can be created by turbulence in simulated clouds. However, they note that an over-dense rim like those found around many of the observed shells is difficult to explain without a driving mechanism providing the momentum to entrain gas.

Accretion-driven winds provide the most likely driving mechanism for the shells. \citet{Offner15} show that a spherical stellar wind with a sufficiently high mass-loss rate can reproduce the shells observed in Perseus by \citet{Arce11}. Below, we compare the winds needed to reproduce the shells in Orion A to winds from intermediate-mass main-sequence stars.

\subsubsection{Wind mass-loss Rates and Energy Injection Rates}\label{sec:wind_energy}
For the following calculations, we assume the shells are driven by winds. Following \cite{Arce11}, we assume that the winds conserve momentum, the wind velocity $v_w$ is 200~\kms, and the duration of the wind $t_w$ is 1~Myr. These values are based on the typical escape velocity of intermediate-mass stars and the approximate age of Class II/III pre-main sequence stars. The wind mass-loss rate that drives a shell with momentum $P_{\rm shell}$ is

\begin{equation}\label{eq:mdot}
\dot{m}_w = \frac{P_{\rm shell}}{v_{\rm w} t_{\rm w}} = 5 \times 10^{-9} M_\odot \rm{yr}^{-1} \frac{P_{\rm{shell}}}{\emph{M}_\odot~\rm{km}~\rm{s}^{-1}}
\end{equation}

The wind mass-loss rates of the shells are reported in Table~\ref{tab:physics}. The rates range from $10^{-8}$ to $8\times 10^{-6}$~\Mdot. These rates are similar to those required by \citet{Offner15} to simulate the types of shells found in Perseus. As noted by \cite{Offner15}, these mass-loss rates are 2-3 orders of magnitudes larger than predicted by theoretical models of line-driven winds from main-sequence B stars \citep[e.g.][Figure 3]{Smith14a}. The discrepancy between the wind mass-loss rates needed to produce the observed shells and the mass-loss rates predicted for line-driven winds from the B and later-type stars present inside the shells shows more modeling of intermediate-mass stellar winds is needed. \cite{Offner15} suggest that periodic wind enhancements due to short term increases in stellar activity or accretion could produce variable mass-loss rates. In this scenario, shells are produced during a short period ($\approx 0.1$ Myr) of enhanced mass-loss while the stars spend most of their lives at the lower mass-loss rates predicted by models. In such a burst, the mass-loss rate would need to increase by an order of magnitude over that estimated by Equation~\ref{eq:mdot}. 

Following \cite{Arce11} and \cite{Li15}, we estimate the wind energy injection rate with Equation 3.7 of \cite{McKee89}:

\begin{equation}
\dot{E}_w = \frac{1}{2} \dot{m}_w v_w \sigma_{\rm 3D}
\end{equation} 

where $v_w~=~200$~\kms~and $\sigma_{\rm 3D}=2.9$~\kms~(see Section~\ref{sec:cloudenergy}). This calculation assumes that the wind deposits its remaining energy on the cloud after radiative losses when it slows to $\sigma_{\rm 3D}$. The wind energy injection rate is distinct from the shell luminosities discussed in Section~\ref{sec:turb}. The total wind energy injection rate is about 14\% of the total mechanical luminosity in the shells. A similar result was found by \cite{Li15}. The power deficit of winds compared to the shells they are driving is likely due to the longer time over which the energy is distributed. The average shell expansion time (from Table~\ref{tab:shells}) is 17\% of the assumed 1 Myr wind duration time. Without better constraints on $t_w$ (and $v_w$), the wind mass-loss rates and energy injection rates are approximate.

Based on the above rates (see Table~\ref{tab:impact}), wind-blown shells may maintain a significant portion of Orion A's turbulence, especially in the North, South, and L1641N subregions.

\section{Summary and Conclusions}
We identify 42 expanding shells in CO maps of the Orion A giant molecular cloud. The shells range in radius from 0.05 to 0.85~pc and are expanding at 0.8 to 5~\kms. Many of the shells are correlated with dust emission and have candidate driving sources near their centers.

We present all 42 shells in the online journal and detail several in this paper: 
\begin{itemize}
\item A C-shaped CO shell near the Herbig A2-3e star T Ori. This pre-main sequence star powers a stellar wind within an order of magnitude of the mass-loss rate needed to drive the CO shell.
\item Two nested shells around the Herbig B9e star V380 Ori. This star is in a hierarchical quadruple system and is responsible for several Herbig-Haro (HH) objects. The dynamical ages of the HH objects are similar to the expansion time of the shells. The shells and outflows traced by the HH objects may have been launched in an accretion-driven outburst during a dynamical interaction among the multiple stellar components of V380 Ori.
\item A shell centered on the B2 pre-main sequence star LP Ori. The mass-loss rate of LP Ori is 2-3 orders of magnitude lower than the wind necessary to drive the expanding shell. 
\end{itemize}

We compare model shells to the CO position-velocity diagrams to estimate their radius, thickness, expansion velocity, and central velocity. Using the models, we extract the H$_2$ mass and calculate momentum, energy, mechanical luminosity, and momentum injection rate of the expanding shells.

The total kinetic energy of the Orion A shells is comparable to the total energy in outflows compiled from the literature. The combined kinetic energy from shells and outflows is significant compared to the turbulent energy of the cloud. The mechanical luminosity and momentum injection rate of the shells and outflows are enough to counteract turbulent dissipation, suggesting that feedback from low to intermediate mass stars may help explain the observed turbulence and low star formation efficiencies in clouds.

One of the mysteries raised by the discovery of CO shells around intermediate-mass stars is the driving mechanism. If the shells are driven by stellar winds, we find wind mass-loss rates ranging from $10^{-8}$ to $8\times 10^{-6}$~\Mdot. These rates are higher than expected for intermediate-mass line-driven stellar winds by 2-3 orders of magnitude. If shells are driven by winds, they probably represent bursts of mass-loss driven by accretion events rather than a continuous flow. A possible source of additional momentum is the heating and ablation of the molecular cloud by FUV photons. Further study of the powering sources and interiors of these shells is needed to resolve the mechanism that drives them.

Orion A marks the third molecular cloud in which expanding shells have been found after Perseus \citep{Arce11} and Taurus \citep{Li15}. Many of these shells show strong evidence for expansion, correlated infrared nebulosity, and candidate sources. Shells have been found in low-mass (Perseus and Taurus) and high-mass (Orion A) star forming regions around intermediate and low-mass stars and are significant to the energetics of these turbulent molecular clouds. These results strongly suggest that further study of the driving sources, mass-loss process, and cloud impact is needed for this new stellar feedback mechanism.

The CARMA-NRO Orion Survey \citep{Kong18} combines the single-dish data used in this paper with interferometry from the Combined Array for Research in Millimeter-wave Astronomy (CARMA). These combined data provide an unprecedented dynamic range in spatial scale - 0.01 to 10~pc - and offer a factor of 3x better resolution compared to the NRO maps alone. This survey will provide a clearer picture of the impact of feedback on the molecular cloud.

\begin{deluxetable*}{ccccccc}

\tablecaption{\label{tab:shells} Shell Parameters}
\tablehead{
\colhead{Shell}&\colhead{Position}&\colhead{$R$\tablenotemark{a}}&\colhead{$dr$\tablenotemark{a}}&\colhead{$v_{\rm exp}$\tablenotemark{a}} &\colhead{$v_{\rm 0}$\tablenotemark{a}}&\colhead{$t_{\rm exp}$\tablenotemark{b}}\\
&\colhead{$\alpha$(J2000),$\delta$(J2000)}&\colhead{(pc)}&\colhead{(pc)}&\colhead{(km s$^{-1}$)}&\colhead{(km s$^{-1}$)}&\colhead{(Myr)}
}

\startdata
$1$&$5^{\rm h}34^{\rm m}49^{\rm s}.0$/$-4\arcdeg51\arcmin41\arcsec$&$0.180\pm0.020$&$0.180\pm0.020$&$1.30\pm0.30$&$10.00\pm0.30$&$0.14\pm0.03$\\
$2$&$5^{\rm h}34^{\rm m}27^{\rm s}.4$/$-4\arcdeg52\arcmin20\arcsec$&$0.100\pm0.010$&$0.100\pm0.010$&$0.80\pm0.20$&$11.40\pm0.20$&$0.12\pm0.03$\\
$3$&$5^{\rm h}34^{\rm m}33^{\rm s}.9$/$-4\arcdeg54\arcmin3\arcsec$&$0.110\pm0.030$&$0.095\pm0.015$&$1.10\pm0.20$&$11.50\pm0.20$&$0.10\pm0.03$\\
$4$&$5^{\rm h}35^{\rm m}23^{\rm s}.9$/$-4\arcdeg55\arcmin15\arcsec$&$0.130\pm0.020$&$0.200\pm0.050$&$1.50\pm0.50$&$11.50\pm0.50$&$0.08\pm0.03$\\
$5$&$5^{\rm h}34^{\rm m}45^{\rm s}.8$/$-4\arcdeg55\arcmin36\arcsec$&$0.120\pm0.020$&$0.075\pm0.025$&$1.15\pm0.15$&$11.70\pm0.20$&$0.10\pm0.02$\\
$6$&$5^{\rm h}35^{\rm m}24^{\rm s}.8$/$-5\arcdeg2\arcmin10\arcsec$&$0.220\pm0.020$&$0.115\pm0.015$&$2.00\pm1.00$&$8.80\pm0.80$&$0.11\pm0.05$\\
$7$&$5^{\rm h}34^{\rm m}54^{\rm s}.5$/$-5\arcdeg4\arcmin40\arcsec$&$0.170\pm0.010$&$0.200\pm0.050$&$3.00\pm1.00$&$14.25\pm0.75$&$0.06\pm0.02$\\
$8$&$5^{\rm h}35^{\rm m}32^{\rm s}.3$/$-5\arcdeg6\arcmin49\arcsec$&$0.150\pm0.050$&$0.175\pm0.025$&$1.75\pm0.25$&$13.75\pm0.25$&$0.08\pm0.03$\\
$9$&$5^{\rm h}34^{\rm m}5^{\rm s}.1$/$-5\arcdeg11\arcmin10\arcsec$&$0.850\pm0.050$&$0.850\pm0.050$&$2.30\pm0.30$&$8.70\pm0.30$&$0.36\pm0.05$\\
$10$&$5^{\rm h}35^{\rm m}43^{\rm s}.0$/$-5\arcdeg27\arcmin47\arcsec$&$0.350\pm0.020$&$0.200\pm0.050$&$3.00\pm1.00$&$11.00\pm0.50$&$0.11\pm0.04$\\
$11$&$5^{\rm h}35^{\rm m}9^{\rm s}.4$/$-5\arcdeg27\arcmin53\arcsec$&$0.220\pm0.040$&$0.230\pm0.030$&$2.00\pm0.50$&$13.00\pm0.50$&$0.11\pm0.03$\\
$12$&$5^{\rm h}36^{\rm m}39^{\rm s}.7$/$-5\arcdeg28\arcmin41\arcsec$&$0.150\pm0.050$&$0.175\pm0.075$&$3.50\pm0.50$&$7.75\pm0.25$&$0.04\pm0.02$\\
$13$&$5^{\rm h}34^{\rm m}24^{\rm s}.2$/$-5\arcdeg29\arcmin0\arcsec$&$0.350\pm0.050$&$0.250\pm0.050$&$2.00\pm0.50$&$6.50\pm0.50$&$0.17\pm0.05$\\
$14$&$5^{\rm h}35^{\rm m}0^{\rm s}.7$/$-5\arcdeg29\arcmin57\arcsec$&$0.210\pm0.040$&$0.190\pm0.020$&$1.40\pm0.50$&$13.50\pm0.50$&$0.15\pm0.06$\\
$15$&$5^{\rm h}34^{\rm m}53^{\rm s}.1$/$-5\arcdeg30\arcmin58\arcsec$&$0.550\pm0.050$&$0.550\pm0.050$&$1.50\pm0.50$&$7.00\pm1.00$&$0.36\pm0.12$\\
$16$&$5^{\rm h}36^{\rm m}25^{\rm s}.1$/$-5\arcdeg33\arcmin38\arcsec$&$0.180\pm0.020$&$0.125\pm0.025$&$2.00\pm1.00$&$7.50\pm0.50$&$0.09\pm0.05$\\
$17$&$5^{\rm h}34^{\rm m}1^{\rm s}.7$/$-5\arcdeg36\arcmin14\arcsec$&$0.250\pm0.050$&$0.250\pm0.050$&$1.35\pm0.15$&$7.20\pm0.20$&$0.18\pm0.04$\\
$18$&$5^{\rm h}34^{\rm m}30^{\rm s}.3$/$-5\arcdeg37\arcmin5\arcsec$&$0.350\pm0.050$&$0.250\pm0.050$&$1.75\pm0.25$&$9.30\pm0.30$&$0.20\pm0.04$\\
$19$&$5^{\rm h}35^{\rm m}24^{\rm s}.0$/$-5\arcdeg45\arcmin57\arcsec$&$0.650\pm0.050$&$0.550\pm0.050$&$5.00\pm1.00$&$6.00\pm1.00$&$0.13\pm0.03$\\
$20$&$5^{\rm h}35^{\rm m}18^{\rm s}.2$/$-5\arcdeg52\arcmin48\arcsec$&$0.145\pm0.015$&$0.125\pm0.025$&$1.25\pm0.25$&$10.00\pm0.10$&$0.11\pm0.03$\\
$21$&$5^{\rm h}34^{\rm m}34^{\rm s}.4$/$-5\arcdeg57\arcmin22\arcsec$&$0.500\pm0.100$&$0.350\pm0.050$&$3.50\pm0.50$&$7.00\pm1.00$&$0.14\pm0.03$\\
$22$&$5^{\rm h}35^{\rm m}0^{\rm s}.1$/$-5\arcdeg59\arcmin17\arcsec$&$0.550\pm0.050$&$0.190\pm0.040$&$1.80\pm0.30$&$10.10\pm0.30$&$0.30\pm0.06$\\
$23$&$5^{\rm h}36^{\rm m}8^{\rm s}.2$/$-6\arcdeg4\arcmin25\arcsec$&$0.650\pm0.050$&$0.300\pm0.100$&$3.00\pm1.00$&$8.75\pm0.75$&$0.21\pm0.07$\\
$24$&$5^{\rm h}35^{\rm m}36^{\rm s}.0$/$-6\arcdeg5\arcmin14\arcsec$&$0.235\pm0.035$&$0.235\pm0.035$&$2.00\pm1.00$&$7.00\pm1.00$&$0.11\pm0.06$\\
$25$&$5^{\rm h}35^{\rm m}15^{\rm s}.7$/$-6\arcdeg15\arcmin22\arcsec$&$0.275\pm0.025$&$0.150\pm0.050$&$1.50\pm0.50$&$7.55\pm0.25$&$0.18\pm0.06$\\
$26$&$5^{\rm h}36^{\rm m}12^{\rm s}.7$/$-6\arcdeg15\arcmin34\arcsec$&$0.175\pm0.025$&$0.150\pm0.050$&$2.50\pm0.50$&$11.00\pm0.30$&$0.07\pm0.02$\\
$27$&$5^{\rm h}35^{\rm m}57^{\rm s}.9$/$-6\arcdeg19\arcmin52\arcsec$&$0.300\pm0.050$&$0.150\pm0.050$&$1.00\pm0.50$&$5.75\pm0.25$&$0.29\pm0.15$\\
$28$&$5^{\rm h}36^{\rm m}10^{\rm s}.2$/$-6\arcdeg24\arcmin7\arcsec$&$0.300\pm0.050$&$0.250\pm0.050$&$1.70\pm0.50$&$9.50\pm0.20$&$0.17\pm0.06$\\
$29$&$5^{\rm h}36^{\rm m}49^{\rm s}.6$/$-6\arcdeg28\arcmin6\arcsec$&$0.170\pm0.020$&$0.125\pm0.025$&$1.20\pm0.20$&$3.75\pm0.25$&$0.14\pm0.03$\\
$30$&$5^{\rm h}36^{\rm m}28^{\rm s}.9$/$-6\arcdeg28\arcmin10\arcsec$&$0.335\pm0.035$&$0.275\pm0.025$&$1.00\pm0.20$&$8.50\pm0.20$&$0.33\pm0.07$\\
$31$&$5^{\rm h}35^{\rm m}40^{\rm s}.0$/$-6\arcdeg29\arcmin51\arcsec$&$0.300\pm0.050$&$0.315\pm0.035$&$1.65\pm0.15$&$7.75\pm0.25$&$0.18\pm0.03$\\
$32$&$5^{\rm h}35^{\rm m}58^{\rm s}.0$/$-6\arcdeg32\arcmin34\arcsec$&$0.260\pm0.020$&$0.220\pm0.020$&$1.50\pm0.50$&$7.70\pm0.30$&$0.17\pm0.06$\\
$33$&$5^{\rm h}36^{\rm m}54^{\rm s}.3$/$-6\arcdeg32\arcmin55\arcsec$&$0.150\pm0.030$&$0.150\pm0.050$&$2.00\pm0.40$&$10.70\pm0.50$&$0.07\pm0.02$\\
$34$&$5^{\rm h}36^{\rm m}24^{\rm s}.4$/$-6\arcdeg35\arcmin0\arcsec$&$0.290\pm0.030$&$0.250\pm0.050$&$2.10\pm0.30$&$9.60\pm0.30$&$0.14\pm0.02$\\
$35$&$5^{\rm h}36^{\rm m}10^{\rm s}.7$/$-6\arcdeg37\arcmin37\arcsec$&$0.250\pm0.050$&$0.250\pm0.050$&$0.85\pm0.15$&$8.00\pm0.10$&$0.29\pm0.08$\\
$36$&$5^{\rm h}36^{\rm m}40^{\rm s}.4$/$-6\arcdeg38\arcmin11\arcsec$&$0.550\pm0.050$&$0.275\pm0.025$&$1.00\pm0.50$&$8.00\pm0.20$&$0.54\pm0.27$\\
$37$&$5^{\rm h}37^{\rm m}6^{\rm s}.5$/$-6\arcdeg38\arcmin20\arcsec$&$0.350\pm0.050$&$0.250\pm0.050$&$2.50\pm0.70$&$7.50\pm0.50$&$0.14\pm0.04$\\
$38$&$5^{\rm h}35^{\rm m}47^{\rm s}.7$/$-6\arcdeg38\arcmin43\arcsec$&$0.280\pm0.030$&$0.150\pm0.050$&$1.00\pm0.50$&$9.00\pm0.50$&$0.27\pm0.14$\\
$39$&$5^{\rm h}36^{\rm m}21^{\rm s}.4$/$-6\arcdeg42\arcmin18\arcsec$&$0.250\pm0.020$&$0.170\pm0.030$&$3.00\pm1.00$&$12.00\pm1.00$&$0.08\pm0.03$\\
$40$&$5^{\rm h}36^{\rm m}23^{\rm s}.1$/$-6\arcdeg42\arcmin27\arcsec$&$0.050\pm0.025$&$0.100\pm0.025$&$1.50\pm0.50$&$11.10\pm0.30$&$0.03\pm0.02$\\
$41$&$5^{\rm h}36^{\rm m}4^{\rm s}.4$/$-6\arcdeg42\arcmin43\arcsec$&$0.175\pm0.025$&$0.200\pm0.050$&$1.75\pm0.75$&$10.40\pm0.60$&$0.10\pm0.04$\\
$42$&$5^{\rm h}38^{\rm m}24^{\rm s}.0$/$-6\arcdeg45\arcmin52\arcsec$&$0.280\pm0.020$&$0.300\pm0.050$&$1.00\pm0.30$&$7.65\pm0.25$&$0.27\pm0.08$\\
\enddata
\tablenotetext{a}{The parameter uncertainties are visually estimated by comparing models to shell PV diagrams.}
\tablenotetext{b}{$t_{\rm exp} = R/v_{\rm exp}$. The uncertainty in expansion time is given by error propagation.}
\end{deluxetable*}



\begin{deluxetable*}{ccccccc}





\tablecaption{Shell Criteria\label{tab:criteria}}


\tablehead{\colhead{Shell} & \colhead{Channel Maps} & \colhead{IR Nebulosity} & \colhead{Circular Structure} & \colhead{PV Diagram} & \colhead{Candidate Source\tablenotemark{a}} & \colhead{Score}}

\startdata
1 & Y (\co[12]) & Y (3.6/8/24/Dust T) & Y (\co[12]) & Y (\co[12]) & Y (Multiple YSO) & 5 \\
2 & Y (\co[12]) & N & Y (\co[12]) & Y (\co[12]) & N & 3 \\
3 & Y ($^{12/13}$CO) & N & Y ($^{12/13}$CO) & Y (\co[12]) & N & 3 \\
4 & Y ($^{12/13}$CO) & N & Y (\co[12]) & Y (\co[12]) & Y (Multiple YSO) & 4 \\
5 & Y ($^{12/13}$CO) & Y (Dust T) & Y ($^{12/13}$CO) & Y (\co[12]) & Y (Multiple YSO) & 5 \\
6 & Y ($^{12/13}$CO) & Y (3.6/8/24) & Y (\co[12]) & Y (\co[12]) & Y (Multiple YSO) & 5 \\
7 & Y (\co[12]) & Y (3.6/8/24) & Y (\co[12]) & Y (\co[12]) & Y (BD-05 1309/A0 \& Multiple YSO) & 5 \\
8 & Y ($^{12/13}$CO) & Y (Dust T) & N & Y (\co[12]) & Y (Multiple YSO) & 4 \\
9 & Y (\co[12]) & N & N & Y (\co[12]) & Y (Brun 193/F9 \& Multiple YSO) & 3 \\
10 & Y ($^{12/13}$CO) & Y (3.6/8) & Y (\co[13]) & N & Y (T Ori/A3e \& Multiple YSO) & 4 \\
11 & Y ($^{12/13}$CO) & Y (3.6/8/24) & Y (\co[12]) & Y (\co[12]) & Y (LP Ori/B2V \& Multiple YSO) & 5 \\
12 & Y ($^{12/13}$CO) & Y (Dust T) & N & N & Y (Brun 1018/B6V) & 3 \\
13 & Y ($^{12/13}$CO) & Y (Dust T) & Y (\co[13]) & Y (\co[12]) & Y (Multiple YSO) & 5 \\
14 & Y ($^{12/13}$CO) & N & Y (\co[12]) & Y (\co[12]) & Y (HD 36939/B7-8II \& Multiple YSO) & 4 \\
15 & Y ($^{12/13}$CO) & Y (Dust T) & N & N & Y (HD 36939/B7-8II \& YSO) & 3 \\
16 & Y (\co[12]) & Y (Dust T) & Y ($^{12/13}$CO) & N & Y ([MGM2012] 1431/YSO) & 4 \\
17 & Y (\co[12]) & N & N & Y (\co[12]) & Y (HD 36782/F5-6V \& Multiple YSO) & 3 \\
18 & Y (\co[13]) & N & Y (\co[13]) & Y (\co[13]) & Y (Multiple YSO) & 4 \\
19 & Y (\co[12]) & N & N & Y (\co[12]) & Y (BD-05 1322/A6V \& Multiple YSO) & 3 \\
20 & Y (\co[12]) & Y (Dust T) & Y (\co[12]) & N & N & 3 \\
21 & Y ($^{12/13}$CO) & N & N & N & Y (Multiple YSO) & 2 \\
22 & Y (\co[12]) & N  & Y (\co[12]) & N & Y (Brun 508/B9V \& Multiple YSO) & 3 \\
23 & Y ($^{12/13}$CO) & Y (Dust T) & Y (\co[13]) & N & Y (Multiple YSO) & 4 \\
24 & Y (\co[12]) & N & N & N & Y (HD 37078/A2V \& Multiple YSO) & 2 \\
25 & Y ($^{12/13}$CO) & Y (Dust T) & Y (\co[13]) & Y (\co[12]) & Y (BD-06 1236/F9 \& Multiple YSO) & 5 \\
26 & Y ($^{12/13}$CO) & N & Y (\co[12]) & Y & N & 3 \\
27 & Y (\co[12]) & Y (3.6/8/24) & N & N & Y ([MGM2012] 969/YSO) & 3 \\
28 & Y ($^{12/13}$CO) & Y (3.8/8/24) & Y (\co[12]) & Y (\co[12]) & Y (BD-06 1251/F5 \& Multiple YSO) & 5 \\
29 & Y ($^{12/13}$CO) & N & N & Y (\co[12]) & N & 2 \\
30 & Y (\co[13]) & Y (3.6/8/24/Dust T) & Y (\co[13]) & N & Y (V1133 Ori/B9IV/V \& Multiple YSO) & 4 \\
31 & Y ($^{12/13}$CO) & Y (3.6/8/Dust T) & N & Y (\co[12]) & Y ([MGM2012] 871/YSO) & 4 \\
32 & Y (\co[12]) & Y (3.6/8/24/Dust T) & Y (IR) & Y (\co[13]) & Y (Multiple YSO) & 5 \\
33 & Y (\co[12]) & Y (Dust T) & N & Y (\co[12]) & Y (Multiple YSO) & 4 \\
34 & Y (\co[12]) & N & N & Y (\co[12]) & Y (Multiple YSO) & 3 \\
35 & Y ($^{12/13}$CO) & N & Y (\co[13]) & Y (\co[12]) & Y (BD-06 1252/F8) & 4 \\
36 & Y (\co[13]) & N & N & N & Y (Multiple YSO) & 2 \\
37 & Y ($^{12/13}$CO) & Y (Dust T) & Y ($^{12/13}$CO) & Y (\co[12]) & Y (Multiple YSO) & 5 \\
38 & Y (\co[13]) & N & N & N & N & 1 \\
39 & Y (\co[12]) & Y (3.6/8/24) & N & Y (\co[12]) & Y (V380 Ori/B9e \& Multiple YSO) & 4 \\
40 & Y ($^{12/13}$CO) & Y (Dust T/HST) & Y (\co[12]) & Y (\co[12]) & Y (V380 Ori/B9e \& YSO) & 5 \\
41 & Y ($^{12/13}$CO) & N & Y (\co[12]) & Y (\co[12]) & Y (Multiple YSO) & 4 \\
42 & Y ($^{12/13}$CO) & Y (3.6/8) & Y ($^{12/13}$CO) & Y (\co[12]) & Y ([MGM2012] 765/YSO) & 5 \\
\enddata
\tablecomments{Entries with Y indicate the shells which satisfy the criteria listed in Section~\ref{sec:criteria}. We also list the observations in which the criteria is most clearly satisfied, among the the two CO spectral cubes and the ancillary data. The ancillary data are indicated as 3.6~=~IRAC~3.6~$\mu$m, 8~=~IRAC~8~$\mu$m, 24~=~MIPS~24$\mu$m, Dust T = \emph{Herschel}/\emph{Planck} dust temperature map, and HST = \emph{HST WFC2}.}
\tablenotetext{a}{If an OBAF-type star is located inside the projected shell radius, we report it as the candidate source. If not, we report YSOs from the Spitzer Orion Survey of \citet{Megeath12}[MGM2012]. When multiple OBAF-type stars are inside projected shell radius, we report the one most likely to drive the shell, based on a combination of spectral type, projected distance to the shell center, parallax, and radial velocity if reported in \emph{Simbad}.}



\end{deluxetable*}

\begin{deluxetable*}{cccccccc}
\tablecaption{\label{tab:physics} Shell Physics}
\tablehead{
\colhead{Shell}&\colhead{$M_{\rm shell}$}&\colhead{$P_{\rm shell}$}&\colhead{$E_{\rm shell}$}&\colhead{$\dot{E}_{\rm shell}$}&\colhead{$\dot{P}_{\rm shell}$}&\colhead{$\dot{m}_{\rm w}$}&\colhead{$\dot{E}_{\rm w}$}\\
&\colhead{(M$_\odot$)}&\colhead{(M$_\odot$ km s$^{-1}$)}&\colhead{($10^{44}$ erg)}&\colhead{($10^{31}$ erg s$^{-1}$)}&\colhead{($10^{-4}$ M$_\odot$ km s$^{-1}$ yr$^{-1}$)}&\colhead{($10^{-7}$ M$_\odot$ yr$^{-1}$)}&\colhead{($10^{31}$ erg s$^{-1}$)}}
 	
\startdata
$1$&$31~[18, 44]$&$41~[18, 71]$&$5~[2, 11]$&$12~[4, 26]$&$3~[1, 5]$&$2~[0.9, 4]$&$4~[2, 6]$\\
$2$&$16~[10, 22]$&$13~[6, 22]$&$1~[0.4, 2]$&$3~[1, 6]$&$1~[0.5, 2]$&$0.6~[0.3, 1]$&$1~[0.6, 2]$\\
$3$&$21~[13, 31]$&$23~[12, 40]$&$3~[1, 5]$&$8~[4, 17]$&$2~[1, 4]$&$1~[0.6, 2]$&$2~[1, 4]$\\
$4$&$42~[22, 64]$&$63~[22, 127]$&$9~[2, 25]$&$35~[8, 95]$&$7~[3, 15]$&$3~[1, 6]$&$6~[2, 12]$\\
$5$&$15~[9, 24]$&$18~[9, 31]$&$2~[0.9, 4]$&$6~[3, 12]$&$2~[0.9, 3]$&$0.9~[0.5, 2]$&$2~[0.9, 3]$\\
$6$&$46~[9, 120]$&$92~[9, 361]$&$18~[0.9, 108]$&$54~[3, 317]$&$9~[0.9, 34]$&$5~[0.5, 18]$&$8~[0.9, 33]$\\
$7$&$29~[7, 67]$&$87~[14, 268]$&$26~[3, 107]$&$149~[15, 610]$&$16~[2, 48]$&$4~[0.7, 13]$&$8~[1, 25]$\\
$8$&$28~[14, 49]$&$48~[21, 97]$&$8~[3, 19]$&$32~[12, 73]$&$6~[2, 12]$&$2~[1, 5]$&$4~[2, 9]$\\
$9$&$220~[145, 274]$&$505~[290, 714]$&$116~[58, 184]$&$101~[51, 162]$&$14~[8, 20]$&$25~[14, 36]$&$46~[26, 65]$\\
$10$&$125~[81, 169]$&$374~[162, 677]$&$112~[32, 269]$&$310~[90, 748]$&$33~[14, 59]$&$19~[8, 34]$&$34~[15, 62]$\\
$11$&$72~[34, 117]$&$146~[52, 295]$&$29~[8, 74]$&$87~[24, 220]$&$14~[5, 28]$&$7~[3, 15]$&$13~[5, 27]$\\
$12$&$24~[14, 40]$&$86~[41, 159]$&$30~[12, 63]$&$225~[93, 478]$&$20~[10, 38]$&$4~[2, 8]$&$8~[4, 15]$\\
$13$&$31~[9, 71]$&$62~[14, 177]$&$12~[2, 44]$&$23~[4, 81]$&$4~[0.8, 10]$&$3~[0.7, 9]$&$6~[1, 16]$\\
$14$&$15~[6, 26]$&$20~[6, 49]$&$3~[0.5, 9]$&$6~[1, 20]$&$1~[0.4, 3]$&$1~[0.3, 2]$&$2~[0.5, 4]$\\
$15$&$284~[168, 416]$&$426~[168, 832]$&$64~[17, 165]$&$56~[15, 146]$&$12~[5, 23]$&$21~[8, 42]$&$39~[15, 76]$\\
$16$&$10~[5, 14]$&$19~[5, 43]$&$4~[0.5, 13]$&$14~[2, 46]$&$2~[0.6, 5]$&$1~[0.3, 2]$&$2~[0.5, 4]$\\
$17$&$23~[13, 37]$&$31~[16, 56]$&$4~[2, 8]$&$7~[3, 15]$&$2~[0.9, 3]$&$2~[0.8, 3]$&$3~[1, 5]$\\
$18$&$102~[63, 147]$&$178~[95, 294]$&$31~[14, 58]$&$50~[23, 95]$&$9~[5, 15]$&$9~[5, 15]$&$16~[9, 27]$\\
$19$&$299~[179, 429]$&$1494~[718, 2574]$&$743~[285, 1535]$&$1852~[711, 3828]$&$117~[56, 202]$&$75~[36, 129]$&$137~[66, 235]$\\
$20$&$6~[4, 9]$&$8~[4, 13]$&$0.9~[0.4, 2]$&$3~[1, 5]$&$0.7~[0.3, 1]$&$0.4~[0.2, 0.7]$&$0.7~[0.4, 1]$\\
$21$&$24~[10, 58]$&$83~[31, 232]$&$29~[9, 92]$&$66~[21, 209]$&$6~[2, 17]$&$4~[2, 12]$&$8~[3, 21]$\\
$22$&$81~[58, 106]$&$146~[87, 222]$&$26~[13, 46]$&$28~[14, 49]$&$5~[3, 7]$&$7~[4, 11]$&$13~[8, 20]$\\
$23$&$335~[190, 490]$&$1005~[381, 1959]$&$300~[76, 779]$&$448~[113, 1165]$&$47~[18, 92]$&$50~[19, 98]$&$92~[35, 179]$\\
$24$&$101~[47, 172]$&$202~[47, 517]$&$40~[5, 154]$&$111~[13, 425]$&$18~[4, 45]$&$10~[2, 26]$&$18~[4, 47]$\\
$25$&$35~[18, 57]$&$52~[18, 113]$&$8~[2, 23]$&$14~[3, 40]$&$3~[1, 6]$&$3~[0.9, 6]$&$5~[2, 10]$\\
$26$&$36~[34, 63]$&$89~[68, 190]$&$22~[14, 57]$&$102~[63, 262]$&$13~[10, 28]$&$4~[3, 9]$&$8~[6, 17]$\\
$27$&$7~[1, 22]$&$7~[0.6, 34]$&$0.7~[0, 5]$&$0.8~[0, 5]$&$0.2~[0, 1]$&$0.4~[0, 2]$&$0.6~[0.1, 3]$\\
$28$&$101~[46, 173]$&$173~[55, 380]$&$29~[7, 83]$&$54~[12, 152]$&$10~[3, 22]$&$9~[3, 19]$&$16~[5, 35]$\\
$29$&$2~[1, 4]$&$3~[1, 5]$&$0.3~[0.1, 0.7]$&$0.8~[0.3, 2]$&$0.2~[0.1, 0.4]$&$0.1~[0.1, 0.3]$&$0.3~[0.1, 0.5]$\\
$30$&$56~[36, 82]$&$56~[29, 99]$&$6~[2, 12]$&$5~[2, 11]$&$2~[0.9, 3]$&$3~[1, 5]$&$5~[3, 9]$\\
$31$&$135~[101, 191]$&$223~[151, 344]$&$37~[23, 61]$&$65~[40, 110]$&$13~[9, 19]$&$11~[8, 17]$&$20~[14, 31]$\\
$32$&$109~[67, 149]$&$163~[67, 299]$&$24~[7, 59]$&$45~[13, 111]$&$10~[4, 18]$&$8~[3, 15]$&$15~[6, 27]$\\
$33$&$8~[3, 15]$&$17~[6, 36]$&$3~[0.9, 9]$&$14~[4, 37]$&$2~[0.8, 5]$&$0.8~[0.3, 2]$&$2~[0.5, 3]$\\
$34$&$40~[22, 66]$&$85~[40, 159]$&$18~[7, 38]$&$42~[17, 89]$&$6~[3, 12]$&$4~[2, 8]$&$8~[4, 15]$\\
$35$&$71~[38, 111]$&$60~[27, 111]$&$5~[2, 11]$&$6~[2, 12]$&$2~[0.9, 4]$&$3~[1, 6]$&$5~[2, 10]$\\
$36$&$114~[46, 180]$&$114~[23, 271]$&$11~[1, 40]$&$7~[0.7, 24]$&$2~[0.4, 5]$&$6~[1, 14]$&$10~[2, 25]$\\
$37$&$112~[69, 149]$&$280~[124, 477]$&$70~[22, 152]$&$161~[51, 351]$&$20~[9, 35]$&$14~[6, 24]$&$26~[11, 44]$\\
$38$&$39~[17, 69]$&$39~[9, 104]$&$4~[0.4, 15]$&$4~[0.5, 18]$&$1~[0.3, 4]$&$2~[0.4, 5]$&$4~[0.8, 9]$\\
$39$&$21~[4, 61]$&$62~[8, 245]$&$18~[2, 97]$&$72~[6, 379]$&$8~[1, 30]$&$3~[0.4, 12]$&$6~[0.7, 22]$\\
$40$&$3~[1, 7]$&$4~[1, 15]$&$0.6~[0.1, 3]$&$6~[1, 28]$&$1~[0.3, 5]$&$0.2~[0.1, 0.7]$&$0.4~[0.1, 1]$\\
$41$&$21~[4, 52]$&$37~[4, 129]$&$7~[0.4, 32]$&$21~[1, 104]$&$4~[0.4, 13]$&$2~[0.2, 6]$&$3~[0.4, 12]$\\
$42$&$19~[9, 26]$&$19~[6, 34]$&$2~[0.4, 4]$&$2~[0.5, 5]$&$0.7~[0.2, 1]$&$0.9~[0.3, 2]$&$2~[0.5, 3]$\\
\enddata
\tablecomments{Best-fit values are reported with lower and upper limits in brackets. Each best-fit value represents the median of an ensemble of models with the best-fit $R$, $dr$, and $v_{\rm exp}$ over the full range in $v_0$ given in Table~\ref{tab:shells}. The lower and upper limits are also ensemble medians using all lower or upper limits of $R$, $dr$, and $v_{\rm exp}$. See Section~\ref{sec:impact_method}.}

\end{deluxetable*}
\begin{deluxetable*}{cccc|cccc|ccc}[htb!]
\tablecaption{\label{tab:impact} Impact of Shells and Outflows}

\tablehead{
\colhead{Subregion}&\colhead{$E_{\rm shells}$\tablenotemark{a}} &\colhead{$E_{\rm out}$\tablenotemark{b}} &\colhead{$E_{\rm turb}$\tablenotemark{c}}& \colhead{$\dot E_{\rm shells}$\tablenotemark{a}} &\colhead{$\dot E_{\rm w}$\tablenotemark{d}}&\colhead{$\dot E_{\rm out}$\tablenotemark{b}}&\colhead{$\dot E_{\rm turb}$\tablenotemark{c}} & \colhead{$\dot P_{\rm shells}$\tablenotemark{a}} & \colhead{$\dot P_{\rm out}$\tablenotemark{b}} & \colhead{$\dot P_{\rm turb}$\tablenotemark{c}}\\
\colhead{Name}&\colhead{($10^{46}$ erg)} &\colhead{ -} &\colhead{-} &\colhead{($10^{33}$ erg s$^{-1}$)}& \colhead{-} & \colhead{-} & \colhead{-} & \colhead{($10^{-3}$ \Msun km s$^{-1}$ yr$^{-1}$)} & \colhead{-} & \colhead{-}
}
\startdata
North&$1.9~[0.7, 4.7]$&$0.68$&$7.8$&$4.0~[1.0, 13.2]$&$0.8~[0.4, 1.6]$&20&2.1&6.0~[1.9, 14.3]&2.0&3.3\\
Central&$2.5~[0.7, 6.4]$&$15$&$20$&$7.2~[2.3, 17.4]$&$1.0~[0.4, 2.0]$&4400&10.6&8.6~[3.6, 16.6]&566&6.9\\
South\tablenotemark{e}&$12~[4.2, 28]$&$-$&$14$&$26.8~[9.7, 60.9]$&$3.0~[1.3, 5.6]$&$-$&2.9&22.1~[10.0, 41.6]&-&3.3\\
L1641N&$2.4~[0.8, 6.2]$&$1.3$&$16$&$5.1~[1.5, 15]$&$1.2~[0.5, 2.5]$&17&4.2&8.2~[3.3, 17.7]&4.6&4.3\\
\hline
Total&$19~[6.4, 45]$&$17$&$58$&$43.1~[14.5, 106]$&$6.1~[2.6, 11.7]$&4437&19.8&44.9~[18.9, 90.2]&573&--
\enddata
\tablenotetext{a}{Shell quantities are given by summing the best-fit values in Table~\ref{tab:physics} corresponding to the shells centered in each subregion (from Figure~\ref{fig:12co_peak_shells}). The lower and upper limits are sums of the lower and upper limits in Table~\ref{tab:physics}.}
\tablenotetext{b}{Outflows are compiled in Section~\ref{sec:outflows}.}
\tablenotetext{c}{Turbulent energies are calculated in Section~\ref{sec:cloudenergy} and injection rates are calculated in Section~\ref{sec:turb}.}
\tablenotetext{d}{Wind energy injection rates are calculated in Section~\ref{sec:wind_energy}.}
\tablenotetext{e}{The shell totals in the South subregion are dominated by two outliers: Shell~19 and Shell~23. Without these two shells, the South subregion totals become $E_{\rm shells} = 1.6~[0.6, 4.9]$, $\dot E_{\rm shells} = 3.8~[1.5, 11]$, $\dot E_{\rm w} = 0.7~[0.3, 1.5]$, and $\dot P_{\rm shells} = 5.7~[2.6, 12.2]$. The total impact from all shells becomes $E_{\rm shells} = 8.4~[2.8, 21.6]$, $\dot E_{\rm shells} = 20.1~[6.7, 53.6]$, $\dot E_{\rm w} = 3.7~[1.6, 7.5]$, and $\dot P_{\rm shells} = 28.5~[11.8, 59.2]$.}

\end{deluxetable*}

\acknowledgments
We thank the anonymous referee for valuable suggestions that improved this paper. JRF was funded by NSF grant AST-1311825 to HGA. This project was also partly funded by AST-1140063. We thank Shuri Oyamada, Sachiko K. Okumura, Yumiko Urasawa, Ryohei Nishi, Kazuhito Dobashi, Tomomi Shimoikura, Takeshi Tsukagoshi, Yoshihiro Tanabe, Takashi Tsukagoshi, and Munetake Momose for their valuable contribution to the 45-m observations. We thank Christopher Beaumont for providing the expanding shell model. We thank Maria Jose Maureira, Mario Tafalla, Crystal Brogan, and Tom Megeath for useful discussions that improved this paper. The Nobeyama 45-m radio telescope is operated by Nobeyama Radio Observatory, a branch of National Astronomical Observatory of Japan. This research has made use of observations made with the NASA/ESA Hubble Space Telescope, and obtained from the Hubble Legacy Archive, which is a collaboration between the Space Telescope Science Institute (STScI/NASA), the Space Telescope European Coordinating Facility (ST-ECF/ESA) and the Canadian Astronomy Data Centre (CADC/NRC/CSA). This research has made use of the NASA/IPAC Infrared Science Archive, which is operated by the Jet Propulsion Laboratory, California Institute of Technology, under contract with the National Aeronautics and Space Administration. This research has made use of the Vizier catalogue access tool \citep{Ochsenbein00} and Simbad database \citep{Wenger00} operated at CDS, Strasbourg, France.

\software{Astropy \citep{Astropy-Collaboration13}, Numpy \citep{numpy}, spectral-cube \citep{Robitaille16}, pvextractor \citep{Ginsburg16}, APLpy \cite{Robitaille12}, Matplotlib \citep{matplotlib}}
\facilities{No:45m, IRSA, MAST, CDS}

\clearpage
\bibliographystyle{aasjournal}
\bibliography{all.bib}

\end{document}